\documentclass{elsart}

 \usepackage{graphicx}

\usepackage{amssymb}

\begin{document}

\begin{frontmatter}



\title{Theory of competition between fusion and quasi-fission in a heavy fusing system}


\author[1,2]{Alexis Diaz-Torres}
\ead{alexis.diaz-torres@anu.edu.au}
\address[1]{Department of Nuclear Physics, Research School of Physical 
Sciences and Engineering, Australian National University, Canberra, 
ACT 0200, Australia}
\address[2]{Institut f\"ur Theoretische Physik der
Johann Wolfgang Goethe--Universit\"at Frankfurt, Max von Laue Str. 1,
D--60438 Frankfurt am Main, Germany}

\begin{abstract}
A theory of the competition between fusion and quasi-fission in a
heavy fusing system is proposed, which is based on a master equation and the two-center
shell model. Fusion and quasi-fission arise from a diffusion process in an ensemble of nuclear shapes, 
each of which evolves towards the thermal equilibrium. The master equation 
describes the diffusion of the nuclear shapes due to quantum and thermal fluctuations. 
Other crucial physical effects like dissipation, ground-state shell effects, diabatic effects and 
rotational effects are also incorporated into the theory. The fusing system moves in a dynamical 
(time-dependent) collective potential energy surface which is initially diabatic and gradually becomes adiabatic. The microscopic ingredients of the theory are obtained with a realistic two-center shell model based on Woods-Saxon potentials. 
Numerical calculations for several reactions leading to $^{256}$No are discussed. 
Among other important conclusions, the results indicate that
(i) the diabatic effects play a very important role in the onset of fusion hindrance for heavy systems, and (ii) very asymmetric reactions induced by closed shell nuclei seem to be the best suited to synthesize the heaviest compound nuclei.    
\end{abstract}

\end{frontmatter}


\section{Introduction}

The fusion of massive nuclei is currently a very hot topic in nuclear physics
motivated by the search for new superheavy elements (SHE)
\cite{Hofmann1,Armbruster,Oganessian1}.
Although the periodic table of the nuclei has been successfully extended during the past
few years by the experimental discovery of new SHE \cite{Hofmann1,Armbruster,Oganessian1}, the
understanding of the formation mechanism of these nuclei is still a challenge for theory. 
Very recent experiments \cite{Loveland} at LBNL in Berkeley (USA) with 
$^{48}$Ca + $^{238}$U have not confirmed the production of the $^{283}$112 SHE reported by the 
Dubna group (Russia) \cite{Oganessian1}.
 
In contrast to the fusion of light and medium mass systems, where the fusion cross
sections are determined by overcoming or tunnelling through the external Coulomb barrier
of the
nucleus-nucleus potential ($\textit{the capture process}$ which is rather well understood
within the
coupled channels framework, e.g., see \cite{Dasgupta1} and references therein),
the fusion cross
section for heavy systems can be inhibited by many orders of magnitude
(fusion hindrance \cite{Toeke,Sahm,Shen1,Reisdorf,Quint,Berriman,Hinde1,Chizhov,Sagaidak,Itkis})
due to the competition
between fusion and quasi-fission (reseparation before compound nucleus (CN)
formation) after the capture stage.
The understanding of this competition remains an unsolved problem and is important 
for optimization of the SHE production rate as well as for understanding the new experimental results 
on fusion of heavy nuclei \cite{Berriman,Hinde1,Chizhov,Sagaidak,Itkis}. 
In addition to the capture process and the subsequent evolution of the
combined system from the contact configuration into the CN
($\textit{the stage of CN formation}$),
the cooling process of the excited CN by the emission of light particles and gamma rays
in competition with fission ($\textit{the deexcitation process}$) clearly affects the
formation of the evaporation residues (ER) observed in the experiments. 
The survival against fission is optimized by minimizing the CN excitation energy and angular momentum. This process is broadly described within a statistical decay model of
atomic nuclei \cite{Ignatyuk1}. It is based on the assumption that all nuclear degrees of 
freedom are equilibrated after the formation of the CN and before it begins to decay. 
For an overview of the three stages of a heavy ion fusion reaction leading to the
formation of a heavy ER, we refer to Ref. \cite{Zagrebaev1}.

Most of the theoretical groups involved in SHE research have a similar viewpoint regarding
the description of the capture stage and the deexcitation process, but there is no consensus
\cite{Volkov} on the modelling of the intermediate stage of CN formation. The physics of the 
current theories sometimes contradict each other. For the sake of simplicity, most of them assume that 
the fusing system is in thermal equilibrium. Depending on the main coordinate for fusion, two sorts of models can be distinguished: 

\begin{itemize}

\item In the first type
\cite{Greiner1,Swiatecki,Aguiar,Tokuda,Denisov,Abe0,Aritomo,Zagrebaev2,Shen,Abe1}, the
fusion occurs in practice essentially along the radial coordinate using either adiabatic potential energy 
surface (PES) obtained with Strutinsky's macroscopic-microscopic method \cite{Greiner1,Aritomo} 
and semi-empirical approaches \cite{Denisov,Zagrebaev2} 
or a liquid drop PES \cite{Swiatecki,Aguiar,Tokuda,Abe0,Shen,Abe1}.
The competition between fusion and quasi-fission caused by the effect of the thermal fluctuactions
has only been included in recent models of this type, e.g., \cite{Abe0,Aritomo,Zagrebaev2,Shen,Abe1}.
The theory in \cite{Greiner1} is completely static. It only predicts the so-called cold fusion valleys in 
the PES that are formed with target-projectile combinations of closed shell nuclei. A large number 
of these combinations were succesfully used at GSI (Germany) to synthesize SHE by cold fusion reactions. 
The macroscopic dynamical models \cite{Swiatecki,Aguiar,Tokuda,Abe0,Aritomo,Zagrebaev2,Shen,Abe1} are 
classical ones based on the concept of mean trajectory which is affected by the PES, the inertia and friction tensors as well as by thermal fluctuations. The theory of Langevin trajectories \cite{Abe0} is quite appealing, but it neglects important quantum effects (quantum fluctuations, ground-state shell effects, diabatic effects) and many of its ingredients are obtained in a rather uncertain manner, e.g., hydrodynamical inertia and friction tensors as well as the PES which, 
in the best situations \cite{Aritomo}, is calculated with a two-center shell model (TCSM) \cite{Maruhn} that is inappropriate for compact nuclear shapes in very asymmetric reactions. Of course, the experimental data are not always explained. A generalized Langevin approach including quantum fluctuations has been proposed recently in 
Ref. \cite{Ayik3}.

\item In the second type \cite{DNS1,DNS2,Li,Nasirov} (dinuclear system (DNS) model),
the fusion happens only in the mass asymmetry coordinate
$\eta=(A_1-A_2)/(A_1+A_2)$  where $A_1$ and $A_2$ are the mass numbers of the nuclei.
The DNS nuclei remain at the contact configuration and exchange nucleons until either
all nucleons have been transferred
from the lighter to the heavier fragment (complete fusion), or the DNS decays
before the CN formation (quasi-fission). The model assumes a sudden (double-folding in frozen
density approximation) PES in the radial
coordinate, while the PES behaves adiabatically along the fusion path in the $\eta$
coordinate. Thus the theory avoids including the complex physics for large overlapping nuclei. 
The model is also attractive due to its simplicity and because its ingredients are obtained in a 
clearer manner (defined in terms of the properties of the individual fragments forming the DNS). 
Although it has been successfully used to explain many experimental
ER cross sections, its theoretical foundation (e.g., the peculiarity of the PES) is not clear enough yet. Possible reasons for this could be the diabatic effects \cite{Alexis1} and large values of the microscopical mass parameter with respect to the neck degrees of freedom \cite{Alexis11}. 

\end{itemize}

The motivations for this work have been (i) to reconcile these conflicting theories, and 
(ii) to incorporate the \textit{multi-particle quantum nature} of the fusing system, rather than assuming a continuous macroscopic fluid. 
The present paper is aimed at investigating the intermediate stage of CN 
formation and, in particular, the competition between fusion and quasi-fission. 
In this study we give a quantum-statistical formulation of the problem, which can be considered as 
the theoretical foundation of the exploratory calculations performed in Refs. \cite{Alexis00,Alexis0}.
The study is based on the
following general ideas (which are well established but have up to now not been used in 
combination in any of the current models of fusion): (i) Once the two nuclei are at the
contact point, the system moves in a multidimensional space of collective coordinates,
(ii) this motion is governed by a master equation, and (iii) the nature
of the single-particle (sp) motion is time-dependent, it is initially
diabatic and then
approaches the adiabatic limit due to residual two-body collisions.
As a result of (iii), the system moves in a
time-dependent (dynamical) 
collective PES which is initially diabatic and gradually becomes adiabatic. Moreover, the
system does not adopt a single (well-defined) shape, but as a result of quantum and
thermal fluctuactions a probability distribution of the nuclear shapes
\cite{Moretto0} can be expected on the dynamical PES.
The fusion is described as
a statistically non-equilibrated evolution process of
nuclear shapes, in which each shape evolves towards the thermal
equilibrium. In addition to the diabatic effects and thermal fluctuations already treated in Ref. \cite{Alexis00,Alexis0}, we now include the effect of quantum fluctuations as well as rotation of the 
combined system on the shape fluctuations.
The microscopic ingredients of the theory are obtained with a realistic TCSM based on Woods-Saxon 
potentials recently suggested in Ref. \cite{TCWS}.
One of the merits of this TCSM is that it permits studying very mass-asymmetric reactions. 
The present theory is unique and more realistic than the current fusion theories because it is founded on microscopical grounds and also for the first time incorporates a wide range of physical effects that crucially influence the formation of the heaviest compound nuclei.  

The diabatic sp motion
arises from the coherent character of the coupling between the intrinsic and
collective degrees
of freedom whilst the dissipation is caused by the residual two-body collisions
\cite{Noerenberg2}. The adiabatic limit refers to
the case when the occupation of the sp energy levels obeys an equilibrated
Fermi-distribution
with a finite temperature. At the diabatic limit \cite{TCWS,Noerenberg2}
(elastic nuclear matter), the nucleons
do not always occupy the lowest free sp energy levels as in the adiabatic case
(plastic nuclear matter), but remain in the diabatic states during collective motion
of the system. Fig. 1 illustrates this situation.
In other words, the probability for Landau-Zener promotion of the
nucleons at the avoided crossings of adiabatic sp levels is close to one. 
This approach is realistic in the approach phase of collisions near
the Coulomb barrier where the total excitation energy per nucleon
$E^{*}\gtrsim 0.03$ MeV \cite{Cass,Alexis1}.
This has been supported by time-dependent Hartree-Fock calculations
\cite{Cassing0}. During the transition from the
diabatic to the adiabatic limit, the nuclear matter is elastoplastic like
glycerine. Possible indications of the diabatic sp motion can be extracted from
the pre-compound emission of fast particles \cite{Cass_probe,Nore_probe1,Rhein} and gamma rays
\cite{Noerenberg3} as well as from the
intrinsic excitation energy per nucleon of two quasi-fission fragments \cite{Nore_probe2}
which is expected to be different for each fragment, i.e., the fragments would have
different temperatures.

\begin{figure}
\begin{center}
\includegraphics[width=12.0cm]{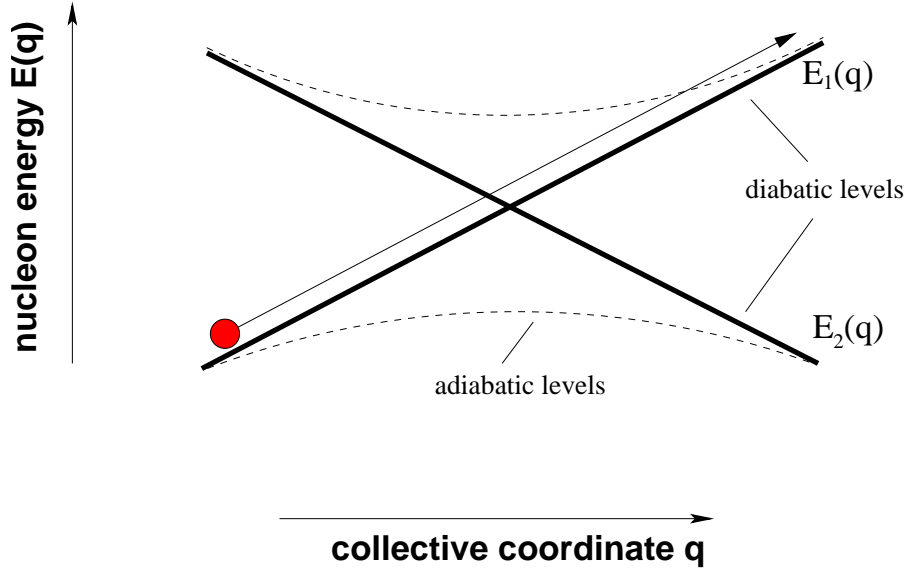}
\end{center}
\caption{Diabatic sp motion at an avoided crossing of two molecular adiabatic sp levels.}
\end{figure}

Experimental evidence to support
the point of view of a quantum-statistical description of the competition between fusion and
quasi-fission can be found, e.g., in Refs. \cite{Reisdorf,Thoennessen}.
Experiments
presented in Ref. \cite{Reisdorf} show that
(i) for a given target-projectile
combination in the entrance channel many complex exit channels, involving large mass and
charge transfers, develop after the passage (inside) of the Coulomb barrier, and (ii) those
complex exit channels reveal that the fragments have different temperatures, i.e.,
the system is not in thermal equilibrium. The experimental study
in Ref. \cite{Thoennessen} clearly indicates that the formation of the CN
occurs in a thermally
non-equilibrated process which depends on the entrance channel.

In section 2 we will present the theoretical formalism in the following order:
(i) the master equation for the evolution of the nuclear shapes is re-derived with
both the coarse-graining procedure used by N\"{o}renberg in Ref. \cite{Noerenberg1}
and methods of the non-equilibrium quantum statistical mechanics
\cite{Zwanzig1,Zwanzig2,Jancel},
(ii) the transition probability between the nuclear shapes is discussed,
(iii) a model of treating the quantum and thermal effects on the shape fluctuations is proposed,
and (iv) a realistic model of calculating the probability
distribution of the nuclear shapes along  with the probability for CN formation
and quasi-fission is suggested. The calculation of other possible $\textit{observables}$ 
like the mass distribution of the quasi-fission fragments as well as the average total and intrinsic excitation energy of the CN is finally discussed.
Here observables mean theoretical quantities that may be verified by direct experiments. 
Of course, the experimental data may contain contributions of many effects that our theory does not include yet. For instance, the experimental mass yield in fusion-fission reactions includes the contribution of fission events, in addition to what we define here as quasi-fission. 
We would like to emphasize that our theoretical mass distribution does not include the fission component (decay of the CN into two fragments), but is limited to all the binary fragmentations which occur after capture and before the CN formation. This is exactly what we call quasi-fission. The separation between the stage of formation of the CN and its fission is justified if the fission time scale is much larger than the formation time of the CN, which is supported by experiments in Ref. \cite{Thoennessen2}.     
In section 3, numerical results for heavy-ion
reactions leading to the $^{256}$No compound nucleus are shown and discussed.
Conclusions are drawn in section 4.

\section{Theoretical formalism}

\subsection{Conceptual details}

The present theory aims at describing the competition between fusion and quasi-fission after capture of the interacting nuclei, translated to a problem of diffusion of nuclear shapes through the multi-dimensional \textit{dynamical} collective potential energy 
landscape. To my knowledge, there are no experimental observations so far, which rule out this scenario. The theory is essentially based on the single-particle model. This may explain and predict the dependence of fusion, after contact of the nuclei, on the shell structure of the interacting nuclei. Owing to the statistical nature of this approach, there is no equation of motion for individual collective coordinates. The theory only describes the evolution in time of an ensemble of nuclear shapes (parametrically defined by a set of collective coordinates) that develop following contact of the interacting nuclei. The basic macroscopic variable will be the nuclear shape, which determines the two-center mean-field in which the nucleons are moving. After capture of the nuclei, the motion of the compact fusing system is expected to be slow (overdamped), as the initial diabatic collective PES practically 
equalizes the total incident energy of the system (small kinetic energy). The usual concept of collective inertia and forces do not appear in the theory explicitly, because the concept of classical trajectory is not applied. Effects similar to the collective inertial effects may be included in the local transition probability rate between the nuclear shapes. This transition probability rate clearly incorporates quantum and thermal effects which drive shape fluctuations.  

The master equation describing the evolution of the nuclear shapes is obtained using the 
time-dependent TCSM representation. The two-body residual interactions are responsible for 
the change of the occupation numbers of the two-center sp orbitals that define the set of many-body channel wave-functions (Slater determinants). The coupling between channels belonging to the same nuclear shape is assumed to be much stronger than the coupling between channels of different nuclear shapes. As a consequence, the Markov limit of the master equation is obtained. This is consistent with our expectation that the evolution in time of the compact nuclear shapes is slow.

The phrase \textit{thermal equilibrium} (local) will only refer to the situation when the 
occupation numbers of the sp orbits at a given nuclear shape obey a Fermi distribution for a finite temperature. Following capture, due to diabatic effects in the entrance phase of the reaction, the sp occupations can be strongly deviated from the Fermi distribution (there is no thermal equilibrium in the system and, strictly speaking, the concept of temperature is meaningless). Diabaticity only produces coherent particle-hole (ph) excitations that contribute to the collective PES. The two-body residual interactions gradually destroy these coherent ph excitations, i.e., the system heats up and the sp occupation numbers evolve in time towards a Fermi distribution for a finite temperature. The evolution in time of the nuclear shapes (a non-equilibrated macroscopic process) occurs in conjunction with the thermalization of each nuclear shape. This scenario will be modelled solving the master equation coupled to the relaxation equation for the sp occupation numbers. We will not include nucleons evaporation. It is important to say that we cannot distinguish between coherent and incoherent ph excitations during the relaxation of the sp occupation numbers. For the sake of simplicity, we will assume that the initial coherent ph excitations turns into incoherent ones (heat) when the thermal equilibrium is reached, i.e., collective modes like surface vibrations of a given nuclear shape are completely damped.
  
\subsection{General aspects}

The total many-body wave-function $|\Psi (t) \rangle$ of the fusing system is
expanded in terms of a set of
many-body $\textit{channel functions}$ $\{ |\Phi_i (t) \rangle\}$ as follows

\begin{equation}
|\Psi (t) \rangle = \sum_i a_i (t)|\Phi_i (t) \rangle, \label{1a}
\end{equation}

which satisfies the time-dependent Schr\"odinger equation

\begin{equation}
i\hbar \frac{\partial |\Psi (t)\rangle}{\partial t} =
\widehat{H} (t) |\Psi (t)\rangle, \label{1b}
\end{equation}

where $\widehat{H} (t)= \widehat{T} + \widehat{V} (t)$ is the total many-body
Hamiltonian. The potential $V (t)$ contains a part which represents the $\textit{two-center
mean field}$ $V_{TCSM}(t)$ and the rest is related to residual nucleon-nucleon interactions
$V_{res}(t)$, i.e., $V(t)=V_{TCSM}(t)+V_{res}(t)$.
The evolution in time of the nuclear shapes is implicitly included by means of the time-dependent
potential $V (t)$. The channel functions $\{ |\Phi_i (t) \rangle\}$ are dynamically defined
by the orthonormalization (e.g., with the Gram-Schmidt method) of the set of Slater
determinants $\{ |\widetilde{\Phi}_i (t)\rangle \}$ (built from the occupied two-center
sp states, not necessarily the lowest orbitals) of all two-center mean field configurations included, i.e.,

\begin{equation}
|\Phi_i (t) \rangle = \sum_j b_{ij} (t)|\widetilde{\Phi}_j (t) \rangle, \label{1bb}
\end{equation}

where the unitary transformation matrix $b_{ij} (t)$ results from the orthonormalization
procedure. The coefficients $a_i (t)$ are the occupation amplitudes
of the individual channels which are assumed to be orthonormal

\begin{equation}
\langle \Phi_j (t)|\Phi_i (t)\rangle=\delta_{ij}, \label{1c}
\end{equation}

satisfying the relation

\begin{equation}
i\hbar \langle \Phi_j (t)|\frac{\partial}{\partial t}\Phi_i (t)\rangle=
\delta_{ij}\langle \Phi_j (t)|\widehat{H}(t)|\Phi_j (t)\rangle. \label{1d}
\end{equation}

Inserting (\ref{1a}) into (\ref{1b}) and projecting on $\langle \Phi_j (t)|$, the
following coupled equations for the amplitudes $a_j (t)$ are obtained

\begin{equation}
i \frac{\partial a_j (t)}{\partial t} =
\sum_{i\neq j} a_i (t) \widetilde{V}_{ji} (t), \label{2a}
\end{equation}
where the non-diagonal matrix elements $\widetilde{V}_{ji} (t) =
\hbar^{-1} \langle \Phi_j (t)|\widehat{H}(t)|\Phi_i (t)\rangle $ yield transitions between
the many-body channels. We now define the
$\textit{density matrix}$

\begin{equation}
\rho_{ji} (t) = a_j (t) a_{i}^{*} (t), \label{3a}
\end{equation}
where the diagonal elements are the time-dependent occupation probabilities of the
individual channels. With (\ref{2a}) and (\ref{3a}),
the Liouville equations for the evolution of $\rho_{ji} (t)$ read as

\begin{equation}
i \frac{\partial \rho_{ji} (t)}{\partial t} =
\sum_{l,k} L_{ji,lk} (t) \rho_{lk} (t), \label{3b}
\end{equation}
where $L_{ji,lk} (t)$ is the Liouville operator defined as

\begin{equation}
L_{ji,lk} (t) = \widetilde{V}_{jl}(t)\delta_{ik} -
\delta_{jl}\widetilde{V}_{ki}(t). \label{3c}
\end{equation}

The operator $L_{ji,lk} (t)$ is a tetradic and behaves like a square matrix in the
superspace $\mathcal{H} \bigotimes \mathcal{H}^{*}$, whereas
$\rho_{lk} = (lk|\rho (t))$ behaves like a vector in that superspace.
$\mathcal{H}$ denotes the Hilbert space of the channel states.
For details about
the algebra of tetradics we refer to Ref. \cite{Zwanzig1}. The total
Hilbert space $\mathcal{H}$ is now divided into subspaces
$\mathcal{H}_{\mu}$ (i.e.,
$\mathcal{H}=\bigoplus \mathcal{H}_{\mu}, \mu = 1,2, \ldots N$). Each subspace
$\mathcal{H}_{\mu}$ defines a $\textit{macrostate}$ with dimension $d_{\mu}$ which
is the number of channels ($\textit{microstates}$) contained in
$\mathcal{H}_{\mu}$. Each macrostate $\mu$ is considered to be uniquely related
to a $\textit{fixed geometry}$ of the two-center mean field
(i.e., a fixed sp spectrum of the
TCSM) that is described by a set of collective ($\textit{shape}$)
coordinates denoted by $\textbf{q}$ (e.g., the relative distance between the two mean field
potentials $\bf{R}$, the so-called fragmentation coordinate $\eta$, etc).
The different many-body channels
belonging to a given $\mu$ are assumed to involve only the Slater determinants
defined by those fixed sp energy levels which are occupied.
It is important to keep in mind that
\emph{each macrostate $\mu$
is assumed to be uniquely associated with a nuclear shape $\textbf{q}$}.
The occupation probability $P_{\mu} (t)$ of the macrostate $\mu$ is defined as

\begin{equation}
P_{\mu} (t)=\sum_{m \mathcal{\epsilon} \mathcal{H}_{\mu}} \rho_{mm} (t).
\label{4}
\end{equation}

These $\textit{macroscopic probabilities}$ $P_{\mu} (t)$ are considered to be the relevant
quantities for describing the evolution of the combined system.
We shall now obtain an equation of motion ($\textit{master equation}$) for these
probabilities.

\subsection{Master equation}

To obtain the master equation for $P_{\mu} (t)$
we should first introduce the
so-called $\textit{coarse-graining operators}$ $C_{\mu}$ \cite{Noerenberg1}
defined as

\begin{equation}
C_{\mu} =d_{\mu}^{-1}\sum_{m,m' \mathcal{\epsilon} \mathcal{H}_{\mu}}
|mm)(m'm'|, \label{4a}
\end{equation}
which act on the vectors of the superspace $\mathcal{H_{\mu}} \bigotimes
\mathcal{H_{\mu}}^{*}$ such as the density matrix (\ref{3a}). These operators are
projectors and fulfil the usual relations $C_{\mu}^{+} = C_{\mu}$ and
$C_{\mu}C_{\nu} = C_{\mu}\delta_{\mu,\nu}$. The operators $C_{\mu}$ act on the
vector $|\rho(t))$ associated with the density matrix taking into account the
definition (\ref{4}) as follows

\begin{equation}
C_{\mu}|\rho (t)) = d_{\mu}^{-1}\sum_{m,m' \mathcal{\epsilon} \mathcal{H}_{\mu}}
|mm)(m'm'|\rho(t)) = d_{\mu}^{-1} P_{\mu} (t)
\sum_{m \mathcal{\epsilon} \mathcal{H}_{\mu}} |mm). \label{4b}
\end{equation}

Using the coarse-graining operators (\ref{4a}), the projection operators
$C=\sum_{\mu} C_{\mu}$ and $Q=1 - C$ can be defined. With the Nakajima-Zwanzig
projection technique described in
Ref. \cite{Zwanzig2} and acting separately with the projectors $C$ and $Q$ on the
Liouville equations (\ref{3b}), the following generalized master equation for the
macroscopic probabilities $P_{\nu} (t)$ is obtained (see Appendix
\ref{Master_Equation})

\begin{equation}
\frac{d P_{\nu} (t)}{d t} = \sum_{\mu \neq \nu}\int_{0}^{t-t_0} d\tau
K_{\nu \mu} (t,\tau)
[d_{\nu}P_{\mu} (t-\tau) - d_{\mu}P_{\nu} (t-\tau)] + G_{\nu} (t,t_0), \label{5}
\end{equation}
where $K_{\nu \mu} (t,\tau)$ is the so-called $\textit{memory kernel}$ and
$G_{\nu} (t,t_0)$ includes the effects on $P_{\nu} (t)$ due to initial
correlations between the channels, i.e., non-diagonal matrix elements $Q|\rho(t_0))$
of the initial density matrix. The quantities
$K_{\nu \mu} (t,\tau)$ (symmetric regarding the indexes $\nu$ and $\mu$) and
$G_{\nu} (t,t_0)$ read as

\begin{eqnarray}
K_{\nu \mu} (t,\tau)=&&-d_{\nu}^{-1}\sum_{n \mathcal{\epsilon} \mathcal{H}_{\nu}}
d_{\mu}^{-1}\sum_{m \mathcal{\epsilon} \mathcal{H}_{\mu}}
(nn|L(t)\exp [-iQ\int_{t-\tau}^{t} dt'L(t')] \nonumber \\
&&\cdot QL(t-\tau)|mm), \label{6a}
\end{eqnarray}

\begin{equation}
G_{\nu} (t,t_0)=-i\sum_{n \mathcal{\epsilon} \mathcal{H}_{\nu}}
(nn|L(t)\exp [-iQ\int_{t_0}^{t} dt'L(t')]\cdot Q|\rho (t_0)). \label{6b}
\end{equation}

In the following, we will
consider that the initial density matrix is diagonal, i.e.
$Q|\rho(t_0))=0$, and thus the inhomogeneity $G_{\nu} (t,t_0)$ in (\ref{5}) is
removed. It is certainly guaranteed if the initial time $t_0$ is fixed before 
the $\textit{decoherence}$ in the process appears as will be discussed below.
Hence, the generalized master equation (\ref{5}) becomes a closed equation
in the macroscopic probabilities $P_{\mu} (t)$.

\textit{Markov limit}. Assuming that the coupling matrix elements $\widetilde{V}_{ji}(t)$
between the channels
are (i) essentially constant during the time interval $\tau$, and
(ii) randomly Gaussian distributed \cite{Nemes},
which is justified if many degrees of freedom are involved in the process 
as is the case in the fusion of a heavy system,
the memory kernel (\ref{6a}) can be approximated as follows
(see Appendix D in Ref. \cite{Noerenberg1})

\begin{equation}
K_{\nu \mu} (t,\tau)\approx 2 \langle \langle |\widetilde{V}_{nm}(t)|^2 \rangle_{\nu}
\rangle_{\mu} \exp [-\pi (\frac{\tau}{\tau_{mem}^{(\nu \mu)}(t)})^2], \label{7}
\end{equation}
where $\langle \ldots \rangle_{\nu}$ and $\langle \ldots \rangle_{\mu}$ denote the mean values
over the channels in the macrostates $\nu$ and $\mu$, respectively, and
$\tau_{mem}^{(\nu \mu)} (t)$ is the so-called $\textit{memory time}$ that determines
the decay of the memory kernel. The averaging coupling and the memory time are defined as
follows

\begin{equation}
\langle \langle |\widetilde{V}_{nm}(t)|^2 \rangle_{\nu}
\rangle_{\mu} =d_{\nu}^{-1}\sum_{n \mathcal{\epsilon} \mathcal{H}_{\nu}}
d_{\mu}^{-1}\sum_{m \mathcal{\epsilon} \mathcal{H}_{\mu}}
|\widetilde{V}_{nm}(t)|^2, \label{7a}
\end{equation}

\begin{equation}
\tau_{mem}^{(\nu \mu)}(t)=\sqrt{2\pi}\{[\sum_{i \mathcal{\epsilon}
\mathcal{H}} (\langle |\widetilde{V}_{ni}(t)|^2 \rangle_{\nu} +
\langle |\widetilde{V}_{mi}(t)|^2 \rangle_{\mu})]\}^{-\frac{1}{2}}. \label{7b}
\end{equation}

The decay of the memory kernel results from the coupling of all the channels which
destroys the $\textit{coherence}$ in the transition between the channels, i.e., the
$\textit{reversibility}$. The coherent character of the sp motion refers to the
$\textit{diabaticity}$ of this motion \cite{Noerenberg2}.
Only during the memory time
$\tau_{mem}^{(\nu \mu)} (t)$ the transition between the macrostates $\mu$ and
$\nu$ is coherent (reversible). It is important to emphasize that the memory time
is the same as the time of coherence. The coherent (diabatic) collective motion is
expected to occur in the approach phase of the reaction at incident energies
slightly above the Coulomb barrier \cite{Cass}.
The memory time was roughly estimated in Ref. \cite{Noerenberg1}
[i.e., $\lesssim (8 \cdot 10^{-21})/A$ s for systems with mass number $A$, e.g.,
$4\cdot 10^{-23}$ s for $A \sim 200$] at energies near the barrier.
If the macroscopic probabilities $P_{\mu} (t)$ are essentially
constant during time intervals of the order of the memory times
$\tau_{mem}^{(\nu \mu)} (t)$, then the generalized master equation (\ref{5}),
with the memory kernel (\ref{7}) and integrating over $\tau$, becomes the following master
equation in the Markov limit (similar to Pauli's master equation (see Ref. \cite{Zwanzig1}))

\begin{equation}
\frac{d P_{\nu} (t)}{d t} = \sum_{\mu \neq \nu} w_{\nu \mu}(t)
[d_{\nu}P_{\mu} (t) - d_{\mu}P_{\nu} (t)], \label{8a}
\end{equation}
where $w_{\nu \mu}(t)$ are the so-called $\textit{transition probability rate}$ (symmetric
with respect to the indexes) defined as

\begin{equation}
w_{\nu \mu}(t)= \tau_{mem}^{(\nu \mu)} (t) \cdot
\langle \langle |\widetilde{V}_{nm}(t)|^2 \rangle_{\nu} \rangle_{\mu}. \label{8b}
\end{equation}

The master equation (\ref{8a}) is valid if the memory times $\tau_{mem}^{(\nu \mu)} (t)$
represent the shortest time scale in the fusion process. The time scale
for the local variations of the macroscopic probabilities $P_{\mu} (t)$ are called
$\textit{relaxation times}$ $\tau_{rel}^{\mu}$,
which are estimated to be of the order of $10^{-21}$ s in
peripheral deep-inelastic heavy ion collisions \cite{Noerenberg1} at the energies considered.
In addition to these time scales, there is a time scale associated with the recurrence of the system close to its initial macroscopic state that is called the 
$\textit{Poincar$\acute{e}$ recurrence time}$
\cite{Jancel}. The Poincar$\acute{e}$ recurrence time has to be the largest one for the
validity of the master equation (\ref{8a}) (see Ref. \cite{Noerenberg1}) and it has been estimated to be several orders of
magnitude larger than the time scales mentioned above in deep-inelastic processes
at the energies discussed \cite{Noerenberg1}.
We would like to stress that the validity conditions of the master
equation (\ref{8a}) in terms of the relation between the different time scales is less restrictive than assuming local thermal equilibrium in each nuclear shape.
The master equation (\ref{8a}) seems to be justified in fusion of heavy nuclei near the Coulomb barrier. 

In the \textit{adiabatic} TCSM the coupling between the channels can be 
expected to be weak, as in this approach the nucleons remain at the lowest free sp orbitals obeying an equilibrated Fermi distribution during the collective motion of the system. 
At very small temperatures, due to Pauli blocking effects, the mean free path of nucleons is expected to be very long. 
Nevertheless, the adiabatic two-center sp states are not used in the present work to define the many-body channels (\ref{1bb}), but the time-dependent two-center sp orbits, i.e., those sp states that are dynamically occupied by the nucleons. Following the entrance phase of the collision, these orbitals are the \textit{diabatic} sp states, and they turn into the \textit{adiabatic} ones during the fusion. My calculations with the realistic diabatic TCSM based on Woods-Saxon potentials \cite{TCWS} indicate that many particle-hole excitations with a considerable energy (particle states up to 10 MeV above the Fermi surface for 
$^{48}$Ca + $^{208}$Pb) can appear following capture, due to diabatic effects. The two-body collisions between a great number of energetic nucleons give rise to transitions between the many-body channels. The life-time of these highly excited sp states [see eq.(\ref{11d})] is indeed very short 
($\sim 10^{-22}$ s) and remains so because the system heats up when the sp occupation numbers gradually evolve to a Fermi distribution for a finite temperature (1-2 MeV). These arguments supported by numerical 
calculations allow me to draw the conclusion that the coupling between the channels is not weak, at least within a given 
macrostate $\mu$ (nuclear shape), but on the contrary strong. As mentioned in subsection 2.1, we will assume that the coupling between channels belonging to different macrostates is much weaker than the coupling between channels within the same macrostate. As discussed in Ref. \cite{Noerenberg1}, this strengthens the validity of the Markov approximation. 
At the moment this is the most practical approach to use in numerical calculations and there is no experimental evidence contradicting it.

Saloner and Weidenm\"uller quantitatively demonstrated in Ref. \cite{Saloner} that (i) in a deeply inelastic collision near the Coulomb barrier, like $^{40}$Ar + $^{232}$Th at an incident energy of $200$ MeV, the strong-coupling condition is fulfilled for the bulk of the interaction region, and (ii) the Markov approximation is indeed justified. For peripheral collisions 
Ayik and N\"orenberg discussed in general in Refs. \cite{Ayik11,Ayik12} the Markov approximation in the limits of weak and strong coupling. In both limits the decay of the memory kernel 
(\ref{6a}) is Gaussian, but the memory time is different \cite{Ayik11}. In the weak-coupling limit the memory time is theoretically determined by the correlation time (see eq. 4.13 in Ref. \cite{Ayik11}) which may be infinite in the overdamped regime of the collective motion. In the strong-coupling limit the memory time (\ref{7b}) corresponds to the decay time of the many-body propagator in (\ref{6a}). In fusion of massive systems the Markov approximation (\ref{8a}) can be expected to be valid because the evolution of the nuclear shapes (where many particles participate) depends on several collective coordinates (deformation, mass and charge asymmetries etc) that may change very slowly. All slowly varying collective coordinates are included in the definition of the nuclear shapes that evolve obeying the master equation (\ref{8a}). To summarize, we
believe that after a short period of coherent motion (memory time or time of
coherence) during the approach phase of the reaction,
the system gradually evolves to a local thermal equilibrium and the probability distribution of the nuclear shapes may be described by the master equation (\ref{8a}). 
Now we will discuss the transition probability rate between the nuclear shapes.  

\subsection{Transition probability rate}

Inserting expressions (\ref{7a}) and (\ref{7b}) into expression (\ref{8b}), the transition
probability rate $w_{\nu \mu}(t)$ can be re-written as

\begin{equation}
w_{\nu \mu}(t)= \frac {\lambda_0^{\nu \mu} (t)} {(d_{\nu} \cdot d_{\mu})^{1/2}}, \label{9a}
\end{equation}
where

\begin{eqnarray}
\lambda_0^{\nu \mu} (t)=&&\sqrt{2\pi} \cdot \sum_{n \mathcal{\epsilon} \mathcal{H}_{\nu}}
\sum_{m \mathcal{\epsilon} \mathcal{H}_{\mu}}|\widetilde{V}_{nm}(t)|^2 \cdot \nonumber \\
&&{\{ \sum_{i \mathcal{\epsilon}
\mathcal{H}} [d_{\mu} \cdot \sum_{n \mathcal{\epsilon} \mathcal{H}_{\nu}}
|\widetilde{V}_{ni}(t)|^2  +
d_{\nu} \cdot \sum_{m \mathcal{\epsilon} \mathcal{H}_{\mu}}|\widetilde{V}_{mi}(t)|^2 ]\}}
^{-\frac{1}{2}}. \label{9b}
\end{eqnarray}

Looking at the master equation (\ref{8a}) we can see that net transition probability rate
$\Lambda_{\nu \mu} (t)$ (not symmetric regarding the indexes) between the macrostates
$\mu $ and $\nu $ can be defined as follows

\begin{equation}
\Lambda_{\nu \mu} (t) = w_{\nu \mu}(t) \cdot d_{\nu} = \lambda_0^{\nu \mu} (t)
\cdot [\frac {d_{\nu}}{d_{\mu}}]^{\frac {1}{2}}, \label{9c}
\end{equation}

\begin{equation}
\Lambda_{\mu \nu} (t) = w_{\nu \mu}(t) \cdot d_{\mu} = \lambda_0^{\mu \nu} (t)
\cdot [\frac {d_{\mu}}{d_{\nu}}]^{\frac {1}{2}}. \label{9d}
\end{equation}

We would like to note that the transition probability rate (\ref{9a}) and
the net transition probability rates (\ref{9c}) and (\ref{9d}) adopt
similar forms as those proposed by Moretto and Sventek in Ref. \cite{Moretto1}
based on phenomenological arguments related to the level density of the system.
This will be demonstrated in the remainder of this subsection.

In the phenomenological expressions in Ref. \cite{Moretto1} the quantity
$\lambda_0^{\nu \mu} (t)$ is a nuclear shape-dependent geometrical parameter and the
dimension of the macrostates is replaced by their level densities $\varrho$. The
dimension of the macrostates ($d_{\nu}$ or $d_{\mu}$) can be expressed in terms
of their entropies following the definition of entropy of the statistical mechanics,
i.e. $S_{\nu} \sim \ln (d_{\nu})$.
Thus $d_{\nu} \sim \exp (S_{\nu}) \sim \varrho (E_{\nu}^{*}(t))$,
being $E_{\nu}^{*}(t)$ the $\textit{total excitation energy}$ of the
macrostate $\nu$. 
Hence, the net transition probability rate, e.g. $\Lambda_{\nu \mu} (t)$, can
be written as

\begin{equation}
\Lambda_{\nu \mu} (t)= \lambda_0^{\nu \mu} (t)
\cdot [\frac {\varrho (E_{\nu}^{*}(t))}{\varrho (E_{\mu}^{*}(t))}]^{\frac {1}{2}} ,\label{10a}
\end{equation}
where the proportionality coefficient between the dimension of the macrostates and the
level density
is assumed to be a constant independent of the macrostate
($d_{\nu}=\Delta E\varrho (E_{\nu}^{*}(t))$, where
$\Delta E=\hbar / \tau_{rel}$ being $\tau_{rel}$ an average of the relaxation times
$\tau_{rel}^{\nu}$ associated with the macroscopic probabilities $P_{\nu}(t)$ discussed above).
The first factor in (\ref{10a}),
$\lambda_0^{\nu \mu} (t)$, essentially includes the $\textit{quantum}$ effects on the evolution of the
macrostates ($\textit{shape fluctuations}$) due to the coupling between the channels,
whereas the second factor is solely related to $\textit{thermal}$ effects.
The quantum effects are actually
not completely separated from the thermal ones in expression (\ref{10a}) owing to the dependence
of $\lambda_0^{\nu \mu} (t)$ on the dimension of the macrostates [see expression (\ref{9b})].
If the Fermi gas formula for level density is used, i.e., $\varrho (x) \sim
\exp (2 \sqrt {ax})$ being $a=A/12$ MeV$^{-1}$ ($A$ is the total mass of the system), then
expression (\ref{10a}) reads

\begin{equation}
\Lambda_{\nu \mu} (t)= \lambda_0^{\nu \mu} (t) \cdot \exp [\sqrt {aE_{\nu}^{*}(t)}-
\sqrt {aE_{\mu}^{*}(t)}]. \label{10b}
\end{equation}

In the limit of a thermally equilibrated collective motion,
the available total excitation energy (shared by the intrinsic and collective
degrees of freedom) can be expressed as
$E_{\nu}^{*}(t) = E - V_{adiab}^{\nu}$, being $E$ the total energy (measured from the
deepest minimum in the potential energy) and $V_{adiab}^{\nu}$ the
so-called adiabatic potential energy. Performing a Taylor expansion of
$\ln \varrho (E - V_{adiab}^{\nu})$ in powers of $V_{adiab}^{\nu}$ and taking contributions up
to the first term, the level density
$\varrho (E_{\nu}^{*}(t)) \approx \varrho (E)\exp (-V_{adiab}^{\nu}/T)$,
where $T$ is the nuclear temperature defined as $T^{-1}=d \ln [\varrho(x)]/dx|_{x=E}$
\cite{Moretto1}.
Inserting this result into expression (\ref{10a}), the well-known expression of the net transition
probability proposed by Moretto and Sventek in Ref. \cite{Moretto1} is obtained, i.e.,
$\Lambda_{\nu \mu} (t) \sim \exp[(V_{adiab}^{\mu} - V_{adiab}^{\nu})/2T]$. Most of the
applications so far \cite{Zagrebaev2,Alexis0,Moretto1,Gippner} have used this approach
in which only the thermal effects on the shape fluctuactions were included.

It is worth mentioning that a semi-microscopical description of the transition probabilities
exits \cite{Ayik1,Adamian1}, but it is valid only if the overlap between the nuclei is small. However, the collective motion, which we are considering, is neither thermally equilibrated nor peripheral
and, therefore, the net transition probabilities may have to be calculated in a different way.
In the next two subsections, we show a model to calculate
the factors $\lambda_0^{\nu \mu} (t)$ and
the total excitation energy $E_{\nu}^{*}(t)$, respectively.

\subsection {Model for calculating the factors $\lambda_0^{\nu \mu} (t)$}

To calculate $\lambda_0^{\nu \mu} (t)$ in expression (\ref{9b}), a modelling
of the non-diagonal coupling between the many-body channels,
e.g., $\widetilde{V}_{nm} (t)$, is needed.
Using the $\textit{diabatic}$ two-center sp basis \cite{TCWS} (which minimizes the radial couplings) and
neglecting the rotational (or Coriolis) couplings (which may be justified if the system is heavy),
this coupling matrix element is essentially given by the coupling between the Slater
determinants $\{ |\widetilde{\Phi}_{i^{'}} (t)\rangle \}$
caused by the residual interaction $V_{res}(t)$, i.e.,

\begin{eqnarray}
\widetilde{V}_{nm} (t)=&&\hbar^{-1} \langle \Phi_n (t)|\widehat{H}(t)|\Phi_m (t)\rangle
\nonumber \\
\approx && \hbar^{-1} \sum_{i^{'},j^{'}} b_{mj^{'}}(t)b_{ni^{'}}^{*}(t)
\langle \widetilde{\Phi}_{i^{'}}(t)|V_{res}(t)|\widetilde{\Phi}_{j^{'}}(t)\rangle, \label{13c}
\end{eqnarray}

provided the transformation matrixes $b_{mj^{'}}(t)$ and $b_{ni^{'}}(t)$ associated with the
macrostates $\mu$ and $\nu$, respectively, are known
(the channel functions $|\Phi_m (t)\rangle$ and $|\Phi_n (t)\rangle$ as well as the set of
Slater determinants $\{ |\widetilde{\Phi}_{j^{'}} (t)\rangle \}$ and
$\{ |\widetilde{\Phi}_{i^{'}} (t)\rangle \}$ belong to the macrostates $\mu$ and $\nu$,
respectively).

We now assume that the two-body residual interaction only depends on the relative
position between the nucleons, i.e., $V_{res}(t)=\sum_{i>j}
v(\textbf{r}_i-\textbf{r}_j)$. The matrix elements among the Slater determinants
in expression (\ref{13c}) will be denoted by $V_{i^{'}j^{'}}(t)$ and they are calculated
as follows

\begin{eqnarray}
V_{i^{'}j^{'}}(t)=(A!)^{-1}&&\sum_{i>j}\mathcal{A}_{(\alpha)}^{\nu}
\mathcal{A}_{(\beta)}^{\mu}\{ \
\langle \varphi_{\alpha_1}(\mathbf{r}_1)|\widetilde{\varphi}_{\beta_1}(\mathbf{r}_1) \rangle
\ldots \nonumber \\
&&\langle \varphi_{\alpha_{i-1}}(\mathbf{r}_{i-1})|\widetilde{\varphi}_{\beta_{i-1}}
(\mathbf{r}_{i-1}) \rangle
\ldots \nonumber \\
&&\langle \varphi_{\alpha_{j-1}}(\mathbf{r}_{j-1})|\widetilde{\varphi}_{\beta_{j-1}}
(\mathbf{r}_{j-1}) \rangle
\cdot \nonumber \\
&&\langle \varphi_{\alpha_{i}}(\mathbf{r}_{i})\varphi_{\alpha_{j}}(\mathbf{r}_{j})
|v(\textbf{r}_i-\textbf{r}_j)|\widetilde{\varphi}_{\beta_{i}}(\mathbf{r}_{i})
\widetilde{\varphi}_{\beta_{j}}(\mathbf{r}_{j}) \rangle
\cdot \nonumber \\
&&\langle \varphi_{\alpha_{i+1}}(\mathbf{r}_{i+1})|\widetilde{\varphi}_{\beta_{i+1}}
(\mathbf{r}_{i+1}) \rangle
\ldots \nonumber \\
&&\langle \varphi_{\alpha_{j+1}}(\mathbf{r}_{j+1})|\widetilde{\varphi}_{\beta_{j+1}}
(\mathbf{r}_{j+1}) \rangle
\ldots \langle \varphi_{\alpha_A}(\mathbf{r}_A)|\widetilde{\varphi}_{\beta_A}
(\mathbf{r}_A) \rangle \ \}, \label{13d}
\end{eqnarray}

where $\varphi_{\alpha}$ and $\widetilde{\varphi}_{\beta}$ are the occupied
diabatic molecular sp states with the set of quantum numbers $\alpha$ and $\beta$,
which build up the Slater determinants
$|\widetilde{\Phi}_{i^{'}} (t)\rangle$ and
$|\widetilde{\Phi}_{j^{'}} (t)\rangle$, respectively. $\mathcal{A}_{(\alpha)}^{\nu}$ and
$\mathcal{A}_{(\beta)}^{\mu}$ refer to the antisymmetrization operators in the macrostates
$\nu$ and $\mu$, respectively, which act on the set of quantum numbers $\alpha$ and $\beta$
of the sp states. These operators are defined as follows, e.g.,
$\mathcal{A}_{(\alpha)}^{\nu}=\sum_P (-1)^{\pi(P)}\ldots$, where $P$ denotes a
permutation of the quantum numbers $\alpha$ among the nucleons
(i.e., $P(1)=\alpha_1, \ldots \ P(A)=\alpha_A$) and $\pi(P)$ corresponds to the number of
inversions (even or odd) of the permutation $P$, i.e., $(-1)^{\pi(P)}=\pm 1$.

\begin{figure}
\begin{center}
\includegraphics[width=12.0cm]{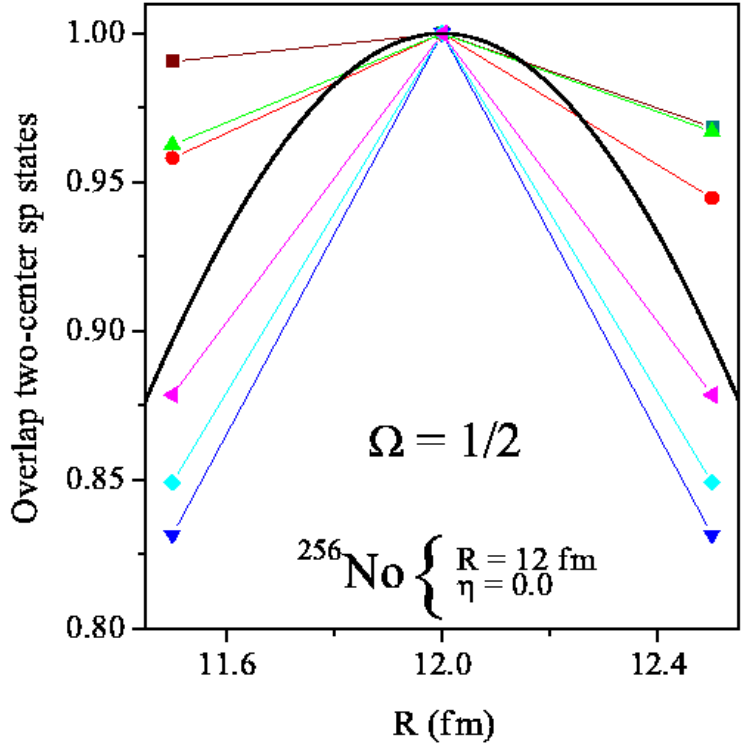}
\end{center}
\caption{(Color online) Diagonal overlap of two-center sp states with the projection $\Omega = 1/2$ of the total angular 
momentum for the symmetric fragmentation of $^{256}$No around $R = 12$ fm. The thick solid curve shows a 
Gaussian fitted by the method of least squares.}
\end{figure}

Using in (\ref{13d}) the antisymmetrization operator $\mathcal{A}_{(\beta)}^{\mu}$ and
considering the diabatic feature of the sp basis, only diagonal terms appear in the
overlap matrix elements of the sp states. Demonstrated in Fig.2 using the TCSM \cite{TCWS} and in
Ref. \cite{Nemes} with the Nilsson-model,
the average of these diagonal overlap matrix elements can be fitted with a local
Gaussian with a variance $\sigma_{\nu}$, e.g.,
$\langle \varphi_{\alpha_1}(\mathbf{r}_1)|\widetilde{\varphi}_{\alpha_1}(\mathbf{r}_1) \rangle
\approx \exp(-(\textbf{q}_{\mu}-\textbf{q}_{\nu})^2/2\sigma_{\nu}^2)$, where
$\textbf{q}_{\nu}$ and $\textbf{q}_{\mu}$ are the collective (shape) coordinates associated
with the macrostates $\nu$ and $\mu$, respectively.
Please note that for reasons of clarity we refer to a Gaussian with a variance,
but it may be a product of Gaussians with different variance where each of them is related
to an individual collective coordinate in the set $\textbf{q}$.
Thus expression (\ref{13d}) becomes

\begin{eqnarray}
V_{i^{'}j^{'}}(t)\approx&&(A!)^{-1}\cdot \exp\{-(A-2)\frac{(\textbf{q}_{\mu}-\textbf{q}_{\nu})^2}
{2\sigma_{\nu}^2}\}\cdot \nonumber \\
&&\sum_{i>j}\mathcal{A}_{(\alpha)}^{\nu}
\{ \
\langle \varphi_{\alpha_{i}}(\mathbf{r}_{i})\varphi_{\alpha_{j}}(\mathbf{r}_{j})
|v(\textbf{r}_i-\textbf{r}_j)|\widetilde{\varphi}_{\alpha_{i}}(\mathbf{r}_{i})
\widetilde{\varphi}_{\alpha_{j}}(\mathbf{r}_{j}) \rangle \}. \label{13e}
\end{eqnarray}

Assuming that $v(\textbf{r}_i-\textbf{r}_j)=\delta(\textbf{r}_i-\textbf{r}_j)$
(zero-range interaction), then

\begin{eqnarray}
\langle \varphi_{\alpha_{i}}(\mathbf{r}_{i})\varphi_{\alpha_{j}}(\mathbf{r}_{j})
|v(\textbf{r}_i-\textbf{r}_j)|\widetilde{\varphi}_{\alpha_{i}}(\mathbf{r}_{i})
\widetilde{\varphi}_{\alpha_{j}}(\mathbf{r}_{j}) \rangle && =
\langle \varphi_{\alpha_{i}}(\mathbf{r}_{i})\varphi_{\alpha_{j}}(\mathbf{r}_{i})
|\widetilde{\varphi}_{\alpha_{i}}(\mathbf{r}_{i})
\widetilde{\varphi}_{\alpha_{j}}(\mathbf{r}_{i}) \rangle \nonumber \\
&&= C_{\alpha_{i},\alpha_{j}}^{i^{'},j^{'}}. \label{13f}
\end{eqnarray}

Denoting by $\overline{C}_{ij}^{i^{'},j^{'}}$ the antisymmetrization of (\ref{13f}) in (\ref{13e})
including the factor $(A!)^{-1}$,
the matrix elements (\ref{13e}) are finally written as

\begin{equation}
V_{i^{'}j^{'}}(t)\approx \exp\{-\frac{(\textbf{q}_{\mu}-\textbf{q}_{\nu})^2}
{2\widetilde{\sigma}_{\nu}^2}\}\cdot \sum_{i>j}\overline{C}_{ij}^{i^{'},j^{'}}, \label{13g}
\end{equation}

where $\widetilde{\sigma}_{\nu}=\sigma_{\nu}/\sqrt{A-2}$. Thus the matrix elements (\ref{13c}) are expressed
as follows

\begin{equation}
\widetilde{V}_{nm} (t)\approx \hbar^{-1} \exp\{-\frac{(\textbf{q}_{\mu}-\textbf{q}_{\nu})^2}
{2\widetilde{\sigma}_{\nu}^2}\}\cdot \mathcal{C}_{nm}, \label{13h}
\end{equation}

where

\begin{equation}
\mathcal{C}_{nm}=\sum_{i^{'},j^{'}} b_{mj^{'}}(t)b_{ni^{'}}^{*}(t)
\ (\sum_{i>j}\overline{C}_{ij}^{i^{'},j^{'}}). \label{13i}
\end{equation}

If we now use expression (\ref{13h}) in expression (\ref{9b}), the quantities
$\lambda_0^{\nu \mu} (t)$ can be written as follows

\begin{equation}
\lambda_0^{\nu \mu} (t) \approx \kappa_0^{\nu \mu}(t)\cdot
\exp\{-\frac{(\textbf{q}_{\mu}-\textbf{q}_{\nu})^2}
{\widetilde{\sigma}_{\nu}^2}\}, \label{13j}
\end{equation}

where

\begin{eqnarray}
&&\kappa_0^{\nu \mu}(t)=\sqrt{2\pi} \hbar^{-1} \cdot \sum_{n \mathcal{\epsilon} \mathcal{H}_{\nu}}
\sum_{m \mathcal{\epsilon} \mathcal{H}_{\mu}}|\mathcal{C}_{nm}(t)|^2 \cdot \nonumber \\
&&\{ \sum_{l \mathcal{\epsilon}
\mathcal{H}} [d_{\mu} \cdot \sum_{n \mathcal{\epsilon} \mathcal{H}_{\nu}}
e^{-\frac{(\textbf{q}_{l}-\textbf{q}_{\nu})^2}{\widetilde{\sigma}_{\nu}^2}} \cdot |\mathcal{C}_{nl}(t)|^2  +
d_{\nu} \cdot \sum_{m \mathcal{\epsilon} \mathcal{H}_{\mu}}
e^{-\frac{(\textbf{q}_{l}-\textbf{q}_{\mu})^2}{\widetilde{\sigma}_{\mu}^2}} \cdot |\mathcal{C}_{ml}(t)|^2 ]\}^
{-\frac{1}{2}}. \nonumber \\ \label{13k}
\end{eqnarray}

The calculation of the dimensions $d_{\nu}$ and $d_{\mu}$ in terms of the level density
was discussed in the previous subsection.

\subsection {Model for calculating the total excitation energy $E_{\nu}^{*}(t)$}

The available total excitation energy $E_{\nu}^{*}(t) \geq 0$ in the rotating body-fixed
reference frame, which can be distributed between
the intrinsic and the remaining collective degrees of freedom, is calculated as follows
\begin{equation}
E_{\nu}^{*}(t)=\mathit{Maximum}\ \{ \epsilon_{\nu}^{*}(t),\ E_{c.m.}-
\frac{\hbar^2 J(J+1)}{2\Theta_{\nu}} - V_{dyn}^{\nu}(t) \},
\label{11_a0}
\end{equation}
where $\epsilon_{\nu}^{*}(t)$ is the $\textit{intrinsic}$
excitation energy, $E_{c.m.}$ is the total incident energy in the center of mass reference frame,
$J$ is the total angular momentum of the system
with a local moment of inertia $\Theta_{\nu}$, and $V_{dyn}^{\nu}(t)$ is the
$\textit{dynamical}$ collective PES defined as
\begin{equation}
V_{dyn}^{\nu}(t)=V_{adiab}^{\nu} (\epsilon_{\nu}^{*}(t)) + \Delta V_{diab}^{\nu}(t),
\label{11_a1}
\end{equation}
where $V_{adiab}^{\nu}$ is the adiabatic PES which
is the sum of the liquid drop energy and the microscopic shell and pairing corrections
obtained with Strutinsky's method.
The so-called $\textit{diabatic contribution}$ $\Delta V_{diab}^{\nu}(t)$
is initially maximal, but
gradually decreases when the sp occupation numbers approach the equilibrated
Fermi distribution. The dynamical PES (\ref{11_a1}) describes a continuous transition
from the initial diabatic potential to the asymptotic adiabatic one.
We would like to stress that the effect of the rotation on the evolution of the
combined system will only be included through expression (\ref{11_a0}). 
This expression clearly shows that in a classically inaccessible macrostate $\nu$ 
[i.e., the second part of (\ref{11_a0}) would be negative] the intrinsic excitation 
energy determines $E_{\nu}^{*}(t)$. 
In the remaining
part of this subsection, we discuss the calculation of the diabatic contribution
$\Delta V_{diab}^{\nu}(t)$ and the intrinsic excitation energy $\epsilon_{\nu}^{*}(t)$.

\textit{Diabatic contribution}. According to the dissipative diabatic dynamics (DDD) approach
(see \cite {Noerenberg3}
and references therein), which is consistent with the features of the collective motion
we aim to describe, the
dissipation arises from the decay of the diabatic contribution $\Delta V_{diab}^{\nu} (t)$
due to the residual
two-body interactions (neglecting the effect of the rotational couplings).
The diabatic contribution $\Delta V_{diab}^{\nu}(t)$ results
from the incident collective
kinetic energy primarily stored in the system as a reversible potential energy
(elastic response) during the initial coherent phase of the sp motion. Following this
coherent stage, this amount of energy gradually turns into heat or intrinsic excitation
energy, i.e., a $\textit{thermalization process}$ occurs. Using the adiabatic and diabatic
sp energy levels of the TCSM \cite{TCWS,Lukasiak1},
$\Delta V_{diab}^{\nu}(t)$ can be calculated as follows \cite{Lukasiak1}

\begin{equation}
\Delta V_{diab}^{\nu} (t) = \sum_{\alpha } e_{\alpha}^{\nu, diab} \cdot n_{\alpha}^{\nu, diab}(t) -
\sum_{\alpha^{'}} e_{\alpha^{'}}^{\nu, adiab}\cdot n_{\alpha^{'}}^{\nu, adiab}(e_F,T), \label{11}
\end{equation}
where $\alpha$ and $\alpha^{'}$ denote the quantum numbers of the diabatic and adiabatic levels,
respectively. The diabatic contribution of neutrons and protons is separately calculated.
The adiabatic quantum numbers are contained in the set of the diabatic ones
\cite{Nikitin}. The diabatic sp occupations are indicated by $n_{\alpha}^{\nu, diab}(t)$, while
the adiabatic occupations by $n_{\alpha^{'}}^{\nu, adiab}(e_F,T)$. 
The adiabatic occupations obey a Fermi distribution for a chemical potential $e_F(t)$ 
and a temperature parameter $T(t)$ that only measures the local intrinsic excitation 
energy $\epsilon_{\nu}^{*}(t)$.
Please note that the time dependence of the chemical potential and the temperature parameter is not 
included
in the expressions of this subsection for reasons of clarity. For a given temperature
parameter the chemical potential is obtained by the conservation of the particle number.
The calculation of $T(t)$ will be explained below. Since the diabatic levels solely differ
from the adiabatic levels at some of the avoided crossings of the adiabatic states
\cite{TCWS,Lukasiak1}, an approximation to the time-dependent diabatic
contribution (\ref{11}) can be written as \cite{Alexis1}

\begin{equation}
\Delta V_{diab}^{\nu} (t) \approx \sum_{\alpha } e_{\alpha}^{\nu, diab} \cdot
[n_{\alpha}^{\nu, diab}(t) - n_{\alpha}^{\nu, adiab}(e_F,T)]. \label{11b}
\end{equation}

The time-dependent diabatic occupation numbers $n_{\alpha}^{\nu, diab}(t)$ (actual
occupations within this model) are described with the following relaxation equation
\cite{Cass}

\begin{equation}
\frac {dn_{\alpha}^{\nu, diab}(t)}{dt} = -\tau_{\nu}^{-1}(t) \cdot
[n_{\alpha}^{\nu, diab}(t) - n_{\alpha}^{\nu, adiab}(e_F,T)], \label{11c}
\end{equation}
where $\tau_{\nu}$ is an average relaxation time (in order to conserve the number of particles).
The initial diabatic occupations $n_{\alpha}^{\nu, diab}(t_0)$ are the same for all the macrostates (shapes) $\nu$, which arise from
some equilibrium distribution $n_{\alpha}^{\nu_0, adiab}(T_0)$. 
Here $\nu_0$ denotes a shape of the entrance system before the diabaticity is manifested, e.g., the nuclei are well-separated at their ground-states. The parameter 
$T_0$ will only be used to simulate a diffused distribution of sp occupations around the Fermi surface of the individual nuclei, i.e., the step function is somewhat smeared out. This can happen although the nuclei are cold due to the pairing correlations. 
Since initially in the entrance channel we deal with two separated nuclei, $T_0$ is in general different for each nucleus. Therefore, $T_0$ will be denoted as a two-component parameter, i.e., $T_0=(T_{01},T_{02})$. Realistic values of $T_0$ can be obtained solving the gap equation \cite{Ring} for each isolated nucleus. The relaxation time $\tau_{\nu}$ is defined as

\begin{equation}
\tau_{\nu}^{-1}(t)=\frac{ \sum_{\alpha}n_{\alpha}^{\nu, diab}(t)
\cdot \,\Gamma_{\alpha}^{\nu}(e_F,T) }
{\hbar\,\sum_{\alpha}n_{\alpha}^{\nu, diab}(t) } , \label{eq_2}
\end{equation}
where the widths $\Gamma_{\alpha}^{\nu}$ of the sp levels due to
residual interactions are obtained with the parametrization
given in Ref. \cite{Helmut}

\begin{equation}
\Gamma_{\alpha}^{\nu}=\Gamma_0^{-1}[(e_{\alpha}^{\nu, diab}-e_F)^2+(\pi T)^2] /
(1 + [(e_{\alpha}^{\nu, diab}-e_F)^2+(\pi T)^2]/c^2), \label{11d}
\end{equation}
where $0.03$MeV$^{-1}\leqslant \Gamma_0^{-1} \leqslant 0.061$MeV$^{-1}$
(depending on the nuclear density) and
$15$MeV$\leqslant c \leqslant 30$MeV.

Using expression (\ref{11b}) and equation (\ref{11c}), the diabatic contribution
$\Delta V_{diab}^{\nu}(t)$ satisfies the following differential equation

\begin{equation}
\frac{\partial \Delta V_{diab}^{\nu}(t)}{\partial t} + \frac{1}{\tau_{\nu} (t)}
\Delta V_{diab}^{\nu}(t) = 0, \label{12a}
\end{equation}
whose solution reads as

\begin{equation}
\Delta V_{diab}^{\nu}(t)=\Delta V_{diab}^{\nu}(t_0)\cdot \exp[-\int_{t_0}^{t}
dt'\tau_{\nu}^{-1}(t')]. \label{12b}
\end{equation}

\textit{Intrinsic excitation}. The temperature parameter $T(t)$ is related to the local intrinsic excitation $\epsilon_{\nu}^{*}(t)$ through the Fermi gas expression 
$\epsilon_{\nu}^{*}(t)=a \cdot T(t)^2$, being $a=A/12$ MeV$^{-1}$ ($A$ is the total mass of the system). 
For the calculation of the local intrinsic excitation energy, we follow the dissipation mechanism
of the DDD approach and make the following $\textit{ansatz}$

\begin{equation}
\frac{d \epsilon_{\nu}^{*}(t)}{dt} = -\frac{\partial \Delta V_{diab}^{\nu}(t)}{\partial t}
= \frac{1}{\tau_{\nu} (t)}
\Delta V_{diab}^{\nu}(t), \label{13a}
\end{equation}
if $E_{c.m.} - \frac{\hbar^2 J(J+1)}{2\Theta_{\nu}} - V_{dyn}^{\nu}(t) \geq 0$, otherwise the intrinsic excitation 
$\epsilon_{\nu}^{*}(t)$ remains constant. The system only heats up if it is in a classical region.
The solution of (\ref{13a}) taking into account expression (\ref{12b}) is as follows

\begin{eqnarray}
\epsilon_{\nu}^{*}(t) = \epsilon_{\nu}^{*}(t_0) + \Delta V_{diab}^{\nu}(t_0)\cdot
\int_{t_0}^{t} dt_1 \tau_{\nu}^{-1}(t_1) \cdot \exp[-\int_{t_0}^{t_1}
dt_2 \tau_{\nu}^{-1}(t_2)]. \label{13b}
\end{eqnarray}

The solution (\ref{13b}) is valid following the coherent period of the reaction.
The initial time $t_0$ will therefore be fixed just after
the coherence period (memory time $\tau_{mem}$),
when the master equation (\ref{8a}) also applies.
The initial diabatic contribution $\Delta V_{diab}^{\nu}(t_0)$, calculated with expression (\ref{11b}), arises
from the previous period of coherence (diabaticity) of the sp motion which
started with the well-separated nuclei having some equilibrated Fermi distribution for the
sp occupations.
Since during the coherence period the system does not warm up, the initial intrinsic
excitation energy of the macrostates
$\epsilon_{\nu}^{*}(t_0)$ is equal to $\epsilon_{\nu_0}^{*}(t_0-\tau_{mem})$. 
We will assume that the well-separated colliding nuclei are cold 
[$\epsilon_{\nu_0}^{*}(t_0-\tau_{mem})=0$ MeV].

Since at the initial time $t_0$ some nuclear shapes (members of the ensemble) may be in a classically 
forbidden region, they start heating up later on, at the time $\widetilde{t_0} > t_0$, when the 
condition $E_{c.m.} - \frac{\hbar^2 J(J+1)}{2\Theta_{\nu}} - V_{dyn}^{\nu}(\widetilde{t_0}) = 0$ 
is fulfilled due to the relaxation of the initial diabatic PES. The initial diabatic contribution 
$\Delta V_{diab}^{\nu}(\widetilde{t_0})$, used in expression (\ref{13b}), is then calculated with 
expression (\ref{12b}).  

\subsection{Model for calculating the probability distribution of the nuclear shapes}

The master equation (\ref{8a}), the net transition probability rate (\ref{10b}) along with the
$\lambda_0^{\nu \mu} (t)$ factors (\ref{13j}) and the total excitation energies (\ref{11_a0})
build up a realistic model to study the
probability distribution of the nuclear shapes during the fusion process, i.e., the
calculation of the macroscopic probabilities $P_{\nu} (t)$. 
First we will describe the calculation of the probability for CN 
formation $P_{CN}$ and the
quasi-fission probability $P_{QF}$, which result from the time-dependent probability distribution
of the nuclear shapes $P_{\nu} (t)$. Afterwards we will discuss the calculation of other possible
observables like the mass distribution of the quasi-fission fragments as well as the average total and intrinsic excitation energy of the CN. It is important to keep in mind that $\textit{all these quantities are referred to the rotating body-fixed
reference frame}$, i.e., the fusing system has a constant total angular momentum $J$.

\subsection{The probability for CN formation and quasi-fission,
the mass distribution of the quasi-fission fragments as well as the average total and intrinsic excitation energy of the CN}

The space of the macrostates or nuclear shapes is divided into three regions: (i) region of
compact shapes around the spherical shape of the CN ($\textit{fusion region}$),
(ii) region of separated fragments beyond the Coulomb barrier ($\textit{quasi-fission region}$),
and (iii) region of intermediate shapes which could lead to fusion or quasi-fission ($\textit{competition region}$). For each fragmentation $\eta=(A_1 - A_2)/(A_1 + A_2)$, being $A_1$ and $A_2$ the mass of the nuclei, the fusion radius $R_F$ is defined as the distance from which the shell structure of the spherical CN manifests. Studying the TCSM levels diagram 
\cite{TCWS} we found that at the radius $R_F$ the lowest sp energy level (with $\Omega=1/2$ being $\Omega$ the projection of the sp total angular momentum on the internuclear axis) reaches a minimal value, e.g., see Fig. 3 
where the adiabatic neutron (bottom) and proton (top) levels correlation diagrams for the reaction 
$^{32}$Al + $^{224}$Ac $\to$ $^{256}$No ($\eta = 0.75$) are shown. 
In this example, $R_F$ is about 6 fm. Fig. 4 shows the fusion radius $R_F$ 
(black solid curve), the contact radius $R_C$ (blue dotted curve) and the radius of the adiabatic Coulomb barrier $R_B$ for a central collision (red solid curve) as a function of the entrance channel $\eta$ leading to $^{256}$No. The radius $R_B$ may change as a function of the total angular momentum $J$. In sect. 3 further details are given.

\begin{figure}
\begin{center}
\includegraphics[width=14.0cm]{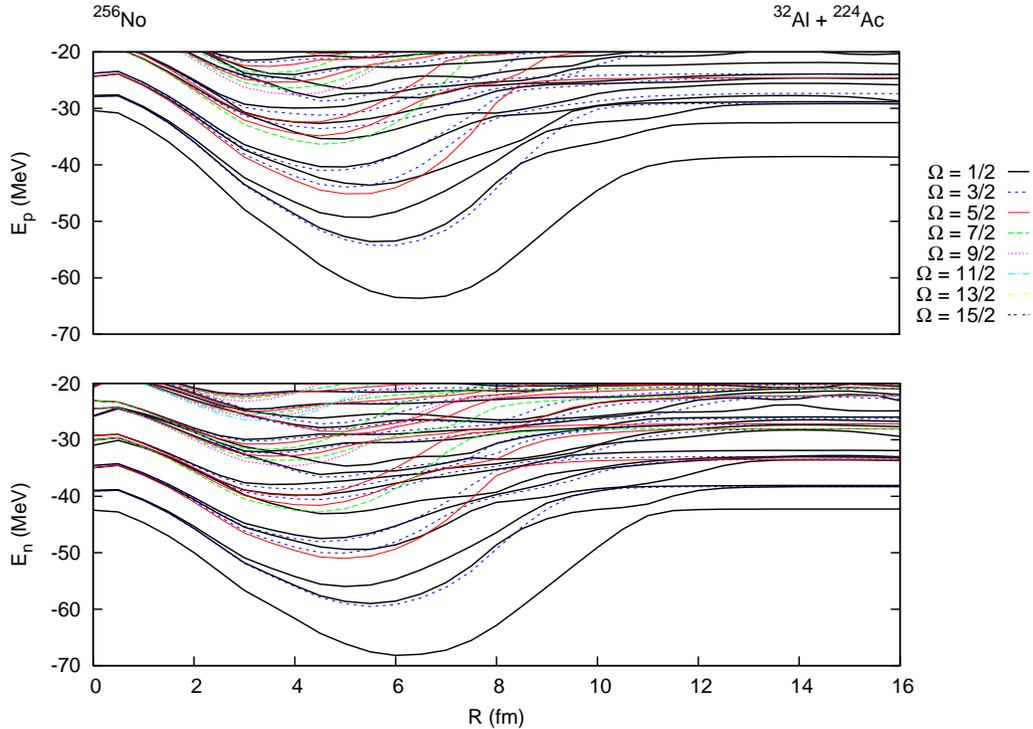}
\end{center}
\caption{(Color online) The lowest adiabatic neutron (bottom) and proton (top) molecular sp levels as a function of the radius 
$R$ between the nuclei for the reaction $^{32}$Al + $^{224}$Ac $\to$ $^{256}$No. Different curves correspond to 
different values of the projection $\Omega$ of the sp total angular momentum on the internuclear axis. See sect. 3 
for further details.}
\end{figure}

\begin{figure}
\begin{center}
\includegraphics[width=14.0cm]{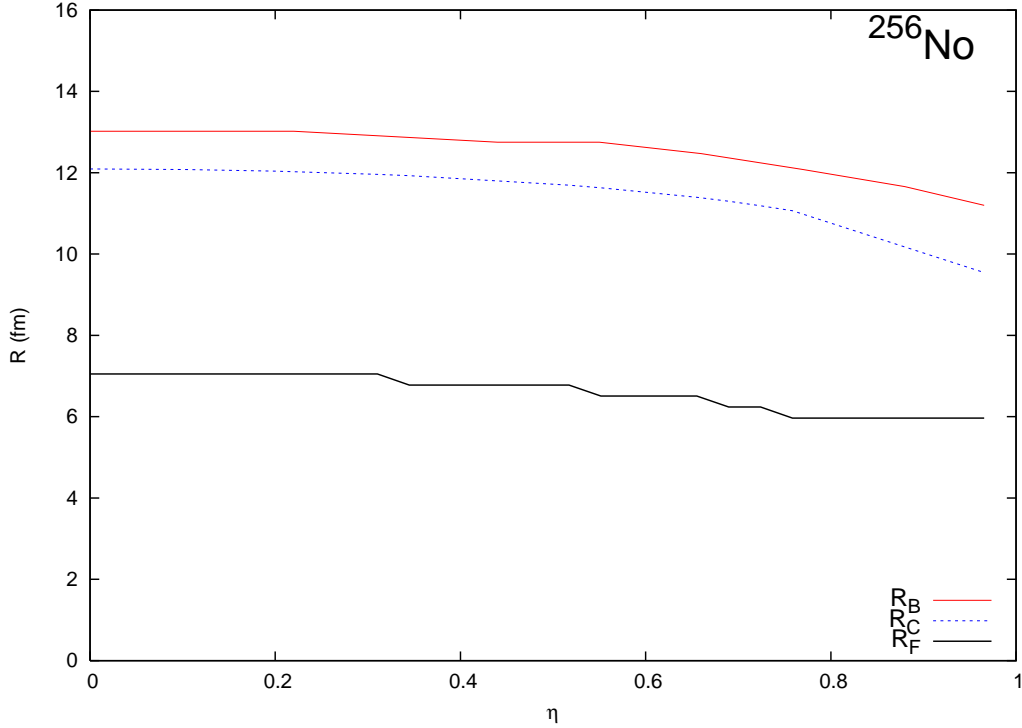}
\end{center}
\caption{(Color online) Radius between the nuclei for an entrance channel $\eta$ 
leading to $^{256}$No, which defines the fusion ($R_F$, black solid curve) and the quasi-fission 
($R_B$, red solid curve) regions. The dotted curve in the competition region denotes 
the contact radius $R_C$. See text and sect. 3 for further details.}
\end{figure}

Initially it is assumed that the system is at the contact configuration (in the competition region inside the Coulomb barrier) with
probability 1. During the evolution of the combined system, the probability
distribution of the nuclear shapes spreads over all the regions and, therefore, the total
occupation probability for each region can be calculated as the sum of the macroscopic
probabilities $P_{\nu} (t)$
belonging to each region. The time-dependent probability for CN formation and
quasi-fission are defined as the sum of the macroscopic probabilities $P_{\nu} (t)$ in the
fusion and quasi-fission regions, respectively. These read as

\begin{equation}
P_{CN} (t) = \sum_{\nu \mathcal{\epsilon} Fus.} P_{\nu} (t), \label{14a}
\end{equation}

\begin{equation}
P_{QF} (t) = \sum_{\nu \mathcal{\epsilon} Qf.} P_{\nu} (t). \label{14b}
\end{equation}

The time scale $\tau_{qf}$ for fusion and quasi-fission is obtained (with a given accuracy) from the condition

\begin{equation}
P_{CN} (\tau_{qf}) + P_{QF} (\tau_{qf}) \approx 1, \label{16}
\end{equation}
i.e., finally the probability for nuclear shapes in the competition region vanishes.
$P_{CN} (\tau_{qf})$ and $P_{QF} (\tau_{qf})$ are the fusion and quasi-fission probabilities,
respectively, observed in the reaction.
The observed mass distribution of the quasi-fission fragments $P_{QF}(\eta,\tau_{qf})$ can be calculated
using $P_{\nu}(\tau_{qf})$ of the nuclear shapes belonging to
the quasi-fission region as follows

\begin{equation}
P_{QF}(\eta,\tau_{qf})=\sum_{\nu \mathcal{\epsilon} Qf.}P_{\nu}(\eta,\tau_{qf}), \label{16a}
\end{equation}

 
Please note that $\nu$ in expression (\ref{16a}) refers to collective
(shape) coordinates other than the fixed mass asymmetry $\eta$, e.g., the radius $R$
between the fragments.
Finally, the average total excitation energy of the CN, i.e. $E_{CN}^{*}(\tau_{qf})$, can be defined as

\begin{equation}
E_{CN}^{*}(\tau_{qf})=\frac{\sum_{\nu \mathcal{\epsilon} Fus.}P_{\nu}(\tau_{qf})
\cdot E_{\nu}^{*}(\tau_{qf})}
{\sum_{\nu \mathcal{\epsilon} Fus.}P_{\nu}(\tau_{qf})}. \label{16c}
\end{equation}

In a similar way, the average intrinsic excitation energy of the CN, i.e. $\epsilon_{CN}^{*}(\tau_{qf})$, can also be calculated with an expression like (\ref{16c}) just replacing 
$E_{\nu}^{*}(\tau_{qf})$ by $\epsilon_{\nu}^{*}(\tau_{qf})$. 

\section{Numerical results}

\subsection{General features and methods}

The basic ingredients of the model are the sp spectra for the different nuclear shapes and the
initial occupation number of the sp levels. The thermal effect on the sp energies can be
neglected at low temperatures ($T \leq 5-6$ MeV) \cite{Bonche},
therefore the sp levels are those of a cold system. The sp spectra for the different nuclear shapes are calculated solving the two-center problem \cite{TCWS} for neutrons and
protons with realistic Woods-Saxon (WS) potentials, while the initial sp occupations are obtained with the diabatic sp motion in the entrance phase of the collision.
The WS potential of the separated spherical nuclei and the spherical CN are obtained with
the global parametrisation given by Soloviev in Ref. \cite{Soloviev}. In a realistic calculation, the potential parameters should be selected in such a way that the experimental sp energies around the Fermi level are reproduced. The two-center
potential of the fusing system is calculated applying the condition of volume conservation
\cite{TCWS}. In order to reduce the computation time in calculating the sp spectra, a small number of 
harmonic oscillator basis states [$l_{max} = 7$ (the number of partial waves in which the potential acts) and 
$n_{max} = 1$ (the number of separable terms in each partial wave)] will be included. 
As collective coordinates we use the radius $R$ between the two WS potentials
(describing the radial motion of the nuclei) and the asymptotic mass asymmetry coordinate
$\eta$ (describing the different fragmentations). We will assume mass/charge equilibrium in the system, i.e., 
the charge asymmetry coordinate $\eta_{Z}= (Z_1 - Z_2)/(Z_1 + Z_2)$ is equal to $\eta$. 

Fig. 5 shows nuclear shapes of the system $^{256}$No as a 
function of $R$ and $\eta$. The shapes with negative $\eta$ values are the same, but the position of the nuclei is reversed. We will use a positive value of $\eta$ for the entrance channel. From $R=0$ up to the contact radius $R_C=1.2\cdot (A_{1}^{1/3}+A_{2}^{1/3})$ fm of the two spherical fragments $A_1=\frac{A}{2}(1+\eta)$ and $A_2=\frac{A}{2}(1-\eta)$, the forms are assumed to be the same as the equipotential shapes defined by the neutron Fermi surface of the fused system. For the sake of simplicity, we have assumed that the nuclear shapes for neutrons coincide with those for protons. The volume of all these shapes is equal to the volume of the spherical $^{256}$No compound nucleus. It is important to notice that our family of nuclear shapes in Fig. 5 does not include shapes with an elongated neck near the contact radius, which are commonly used for describing the fission process. This may not significantly affect our results if the fission valley is mainly reached through compact fused shapes. In a recent work by Ichikawa et al. in Ref. \cite{Moeller} it was shown for cold fusion reactions that a ridge of several MeV, around the contact radius, separates the fusionlike valley from the fission valley. The height and persistence of that ridge for compact shapes may influence the CN formation. Of course, this issue needs further investigation. As mentioned in the introduction, we do not include fusion-fission events in the present calculations. Moreover, we will not include quasi-fission events that may happen through the fission valley.

\begin{figure}
\begin{center}
\includegraphics[width=14.0cm]{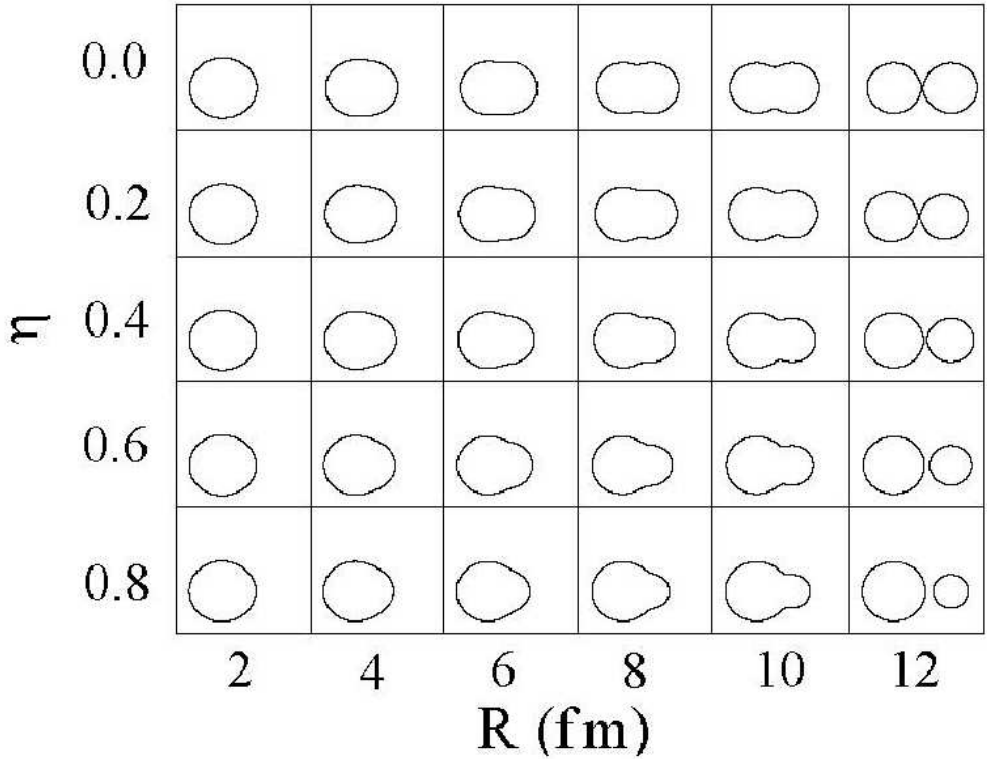}
\end{center}
\caption{Nuclear shapes of the system $^{256}$No as a function of $R$ and $\eta$. See text for further details.}
\end{figure}

Using the axial symmetric shapes like those in Fig. 5, the nuclear part of the liquid drop energy (LDE) in the dynamical collective PES (\ref{11_a1}) is calculated with the Yukawa-plus-exponential method \cite{KNS}. For the separated nuclei the LDE is equal to the Krappe-Nix-Sierk (KNS) potential \cite{KNS}. The KNS potential very well agrees with the recent semi-microscopical potential calculated by Denisov and N\"orenberg in Ref. \cite{Denisovpot} using the formalism of the energy-density functional along with an extended Thomas-Fermi approximation. The shell corrections are obtained with the method recently proposed by the author in Ref. \cite{ShellC}, while the pairing corrections to the PES will be neglected. 
The dependence of the shell corrections on the intrinsic excitation energy $\epsilon_{\nu}^{*}(t)$ is included as usual, i.e., 
by using an exponential damping factor $\exp (-\gamma \epsilon_{\nu}^{*})$ being 
$\gamma=(1+1.3A^{-1/3})/(5.48A^{1/3})$ MeV$^{-1}$ \cite{Ignatyuk2}. 
The moment of inertia of the compact nuclear shapes for the rotation around an axis that is perpendicular to the axial symmetric axis is calculated using the rigid-rotor approximation. 
For the separated fragments, the moment of inertia is that corresponding to the orbital motion, i.e., $\mu R^2$ being $\mu$ the reduced mass for a fragmentation $\eta$. We will study the dependence of the observables on the angular momentum $J$ of the combined system.

Since the diabatic sp excitations occur around the Fermi surface,
the values $\Gamma_0^{-1}=0.061$ MeV$^{-1}$ for half saturation density and $c=20$ MeV
can be used in expression (\ref{11d}) as standard values for calculating the width of the sp levels along with the local relaxation times (\ref{eq_2}). It can be anticipated that the calculation very weakly depends on the parameter $c$, while the dependence on $\Gamma_0^{-1}$ will be studied below. The dependence of the calculation on small values of 
$T_0$ simulating the pairing correlations in the entrance system will not be investigated in the present work, therefore a step function [$T_0 = (0.0,0.0)$ MeV] will be adopted for the initial sp occupations around the Fermi surface of the colliding individual nuclei. 
Besides these parameters involved
in calculating the total excitation energies $E_{\nu}^{*}(t)$,
the first term of the $\lambda_0^{\nu \mu} (t)$ factors (\ref{13j}), i.e. $\kappa_0^{\nu \mu} (t)$,
will be regarded as a constant
parameter $\kappa_0$ that is independent of the macrostates. This approach means
that only the gross feature of the quantum effects on the shape fluctuations will be
included. Experiments related to the quasi-fission of heavy systems with $Z=108$,
which were analyzed \cite{Gippner} with a diffusion model based
on both the master equation (\ref{8a}) and net transition probabilities like those by Moretto
and Sventek discussed in subsect. 2.3, indicate that a realistic value of $\kappa_0$ is of the
order of $10^{22}$ s$^{-1}$ like the magnitude of the factor $\sqrt{2\pi} \hbar^{-1}$
in (\ref{13k}). We will study the dependence of the calculation on $\kappa_0$ values around this 
standard value. 

\begin{figure}
\begin{center}
\includegraphics[width=12.0cm]{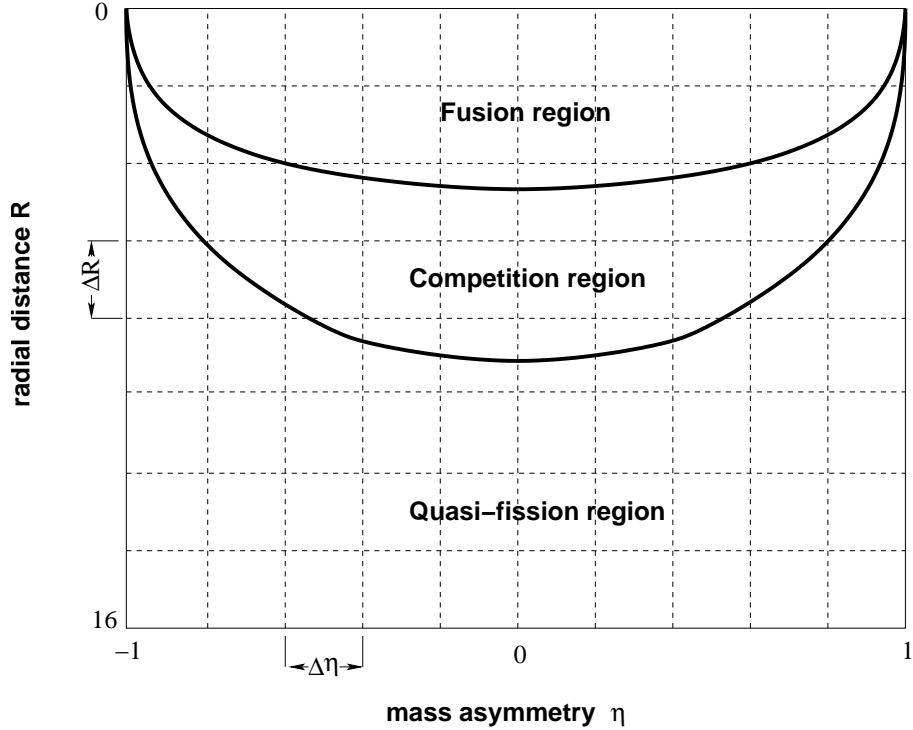}
\end{center}
\caption{Schematic illustration of the 2D-mesh used in the calculation. A node of the mesh is related to a 
nuclear shape. The thick solid curves separate the fusion, competition and quasi-fission regions. 
See text for further details}
\end{figure}

The local variance of the Gaussians 
in (\ref{13j}), i.e. $\widetilde{\sigma}_{R}^{\nu}$ and $\widetilde{\sigma}_{\eta}^{\nu}$, can be obtained by means of the overlap of the TCSM sp wave-functions as shown in Fig. 2. However, in order to simplify the calculation, these variances will be considered as constants ($\widetilde{\sigma}_{R}$ and $\widetilde{\sigma}_{\eta}$) that only depend on the region (fusion, competition and quasi-fission) of the nuclear shapes. Moreover, we will assume that $\widetilde{\sigma}_{R}$ is the same in all regions, whereas $\widetilde{\sigma}_{\eta}$ adopts the same value for overlapping nuclei (fusion and competition regions) and it is essentially zero in the quasi-fission region. For well-separated nuclei, as the shapes belonging to the quasi-fission region, the motion in the mass asymmetry coordinate $\eta$ (nucleon transfer) is expected to be negligible. Unless other values of $\widetilde{\sigma}_{\eta}$ and $\widetilde{\sigma}_{R}$ are pointed out, we will use as standard values in the calculations 
$\widetilde{\sigma}_{\eta} = (0.2,0.2,10^{-6})$ (each component refers to the fusion, the competition and 
the quasi-fission region, respectively) and $\widetilde{\sigma}_{R} = 0.5$ fm.        

The set of equations (\ref{8a}), (\ref{10b}), (\ref{13j}) and (\ref{11_a0}) are solved by successive iterations using a small time step $\Delta t = 10^{-23}$ s. The master equation (\ref{8a}) is solved on a 2D-mesh (Fig. 6) defined by the coordinates $R$ and $\eta$ ($R \leq 16$ fm and $\vert \eta \vert \leq$ 1 with steps $\Delta R$ and $\Delta \eta$ defined below). 
The step values $\Delta R$ and $\Delta \eta$ should be smaller than the Gaussian variances 
$\widetilde{\sigma}_{R}$ and $\widetilde{\sigma}_{\eta}$, respectively, so that the mesh points can be interconnected. As standard values in the calculations below the values $\Delta R = 0.27$ fm and 
$\Delta \eta = 0.1$ will be used. At $\eta=\pm 1$ (spherical CN) only the mesh point with $R=0$ is included because only this point can be physically realized. 
For the sake of simplicity, as initial condition we will assume
that $P_{\nu} (t_0) = \delta_{\nu \nu_0}$ denoting $\nu_0$ the contact configuration of the
entrance system, i.e., the system is well located at the contact configuration.
Hence, we will neglect shape fluctuations that might happen during the capture stage.
This effect may be included using
a Gaussian distribution function for $P_{\nu} (t_0)$ instead of a delta function. 
The boundary conditions are defined in such a way that the initial $P_{\nu}(t_0)$ $\textit{irreversibly}$ flows to the fusion and the quasi-fission regions. All mesh points (shapes) belonging to the fusion and the competition region, respectively, are interconnected. In the quasi-fission region, where the nuclei are well-separated, only the decay of the system along $R$ occurs. The accuracy for expression (\ref{16}) will be about $99$\% .

\begin{figure}
\begin{center}
\includegraphics[width=14.0cm]{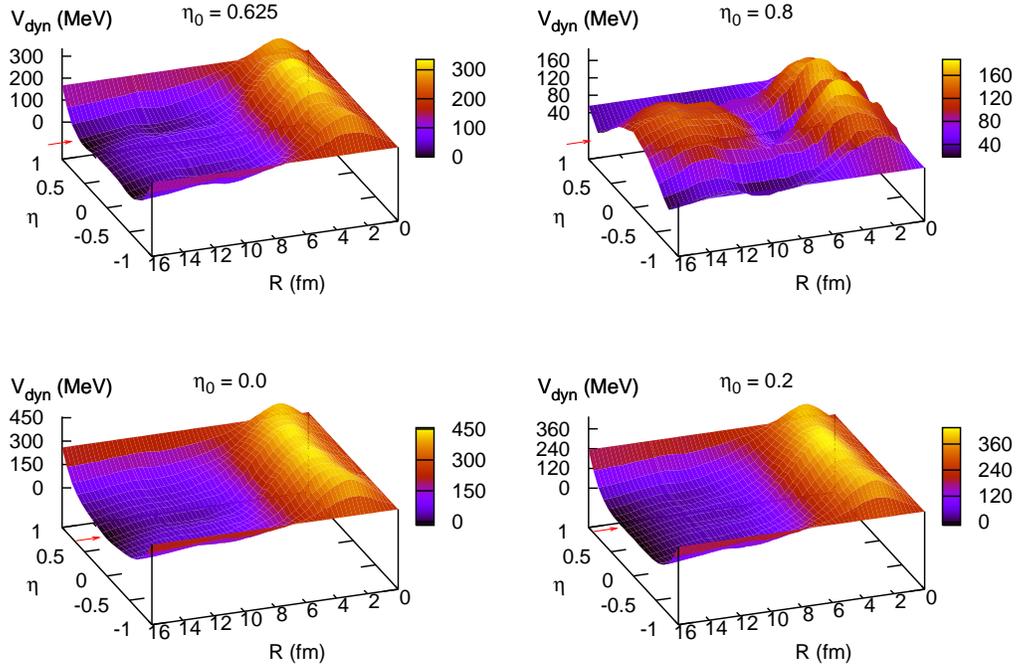}
\end{center}
\caption{(Color online) Entrance diabatic PES as a function of $R$ and $\eta$ for different entrance channel mass asymmetries 
$\eta_0$ (arrow) leading to the system $^{256}$No. See text for further details.}
\end{figure}

\begin{figure}
\begin{center}
\includegraphics[width=14.0cm]{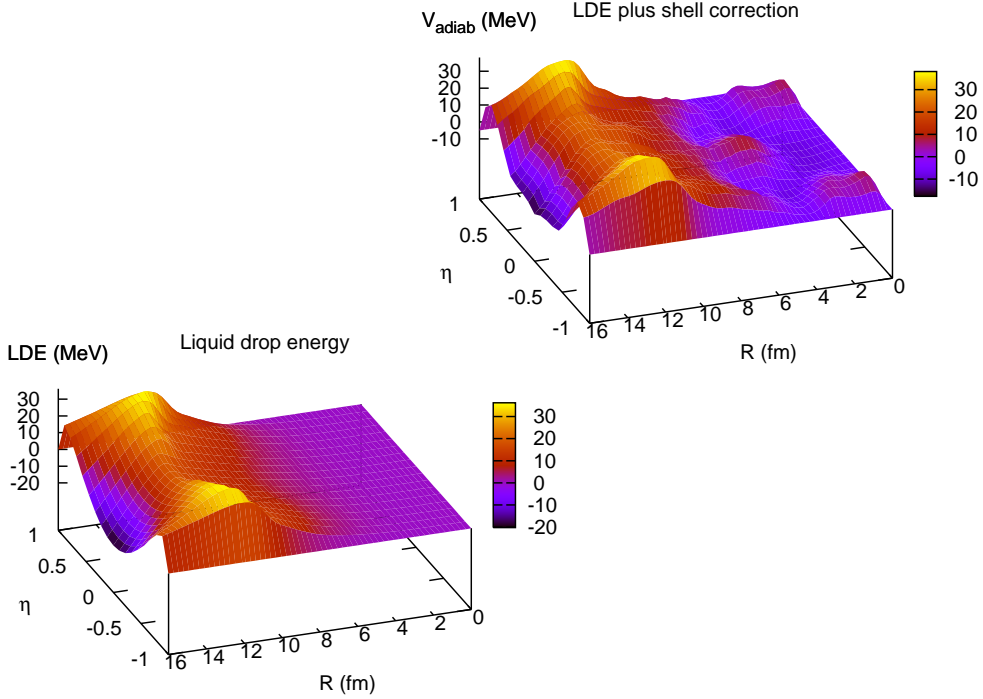}
\end{center}
\caption{(Color online) Adiabatic PES as a function of $R$ and $\eta$ for the cold system $^{256}$No including 
ground-state shell corrections (top) and the macroscopic LDE only (bottom). 
The warm fusing system may reach an asymptotic adiabatic PES between these two 
during the reaction. 
See text for further details.}
\end{figure}

\subsection{Performance of the model}

\subsubsection{Entrance diabatic and asymptotic adiabatic PES}

Fig. 7 shows the entrance diabatic PES as a function of $R$ and $\eta$ for different mass asymmetries $\eta_0$ of the entrance channel leading to the $^{256}$No compound nucleus. The energy is normalized with respect to the LDE of the 
spherical $^{256}$No. 
Fig. 8 (top) shows the adiabatic PES of the cold system (the ground-state shell corrections are included), while at the bottom 
in Fig. 8 one can observe the LDE only, where the quantum shell effects are removed. 
The warm fusing system may reach an asymptotic adiabatic PES between these two PES after a gradual relaxation of the entrance 
diabatic PES of Fig. 7 during the reaction. 
Fig. 9 shows the diabatic capture barrier (black solid curve) for the entrance channels shown in Fig. 7 along with their reference adiabatic barriers including shell corrections (red solid curve) as well as the LDE only (blue dashed curve). The arrow indicates the contact radius $R_C$. In Fig. 10 we show the so-called $\textit{driving potential}$ (defined as the potential energy of the dinuclear system as a function of $\eta$ at the contact radius $R_C$) arising from the entrance diabatic PES of Fig. 7 (curves other than the two lowest ones which are also presented in the small figure inserted, more details appear in Fig. 10) and from the asymptotic adiabatic PES of Fig. 8 [red solid curve (with shell corrections) and black solid curve (LDE only)]. The difference between the entrance diabatic driving potential and the asymptotic adiabatic one is due to the diabatic contribution $\Delta V_{diab}$ 
(\ref{11_a1}). 
From studying Figs. 7-10, we draw the following conclusions:

\begin{itemize}

\item The entrance diabatic PES show a repulsive core at small radii $R$ as well as at large mass asymmetries $\eta$, which decreases with increasing entrance channel 
asymmetry $\eta_0$.
\item The diabatic effects generally cause a capture valley in the PES, which is more pronounced with increasing asymmetry $\eta_0$ of the reaction.
\item The entrance diabatic driving potential largely differs from the asymptotic adiabatic one, so the entrance diabatic PES does not completely justify the PES assumed in the DNS-model \cite{DNS1,DNS2,Li,Nasirov} 
of fusion.
\item The entrance diabatic PES show structures caused by the diabatic sp motion. It is not related to the well-known (static) ground-state shell corrections to the macroscopic LDE, but to the sp motion through the shell structure of the 
different nuclear shapes.
\item First shown by Berdichesky et al. in Ref. \cite{Berdichesky}, the diabatic effects raise the capture barrier (diabatic shift) and slightly reduce the barrier radius, which depends on the entrance channel. This diabatic shift seems to decrease with increasing entrance channel asymmetry $\eta_0$. In some reactions a diabatic screening of the adiabatic potential pocket can occur. Indeed the present results confirm that the diabatic effects in the capture stage of the reaction should play an important role in the onset of fusion hindrance for heavy systems. This will be thoroughly studied in a separate paper.   

\end{itemize}     

\begin{figure}
\begin{center}
\includegraphics[width=14.0cm]{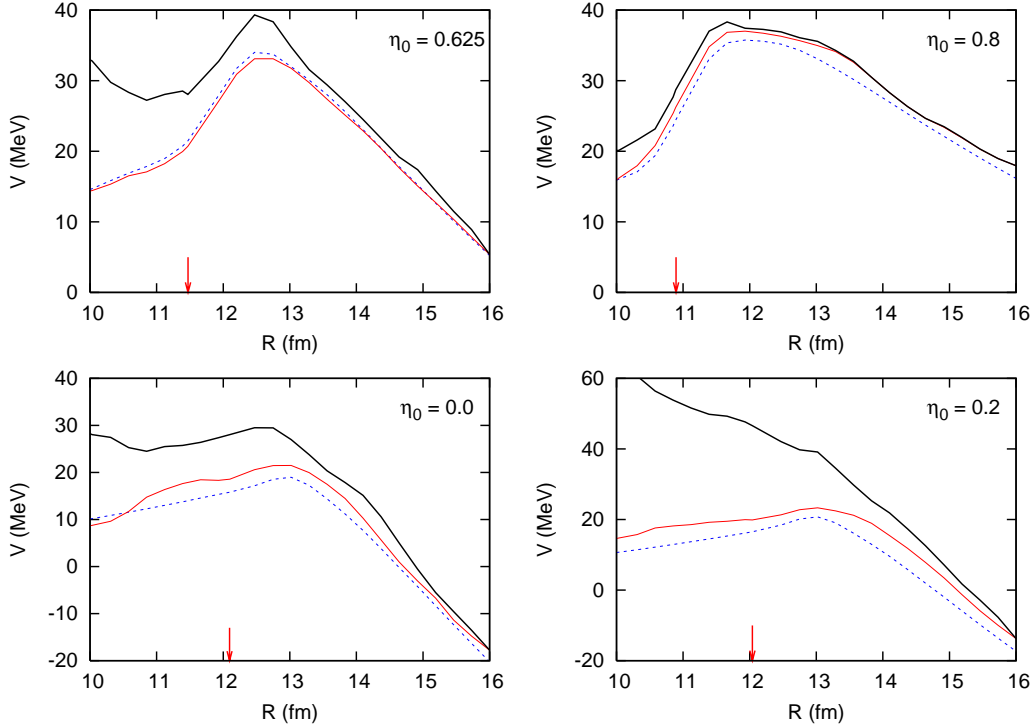}
\end{center}
\caption{(Color online) Entrance diabatic capture barrier (black solid curve) for the reactions shown in Fig. 7 along 
with their reference adiabatic barriers (red solid curve) and the LDE only (blue dashed curve). The arrow 
indicates the contact radius. See text for further details.}
\end{figure}

\begin{figure}
\begin{center}
\includegraphics[width=14.0cm]{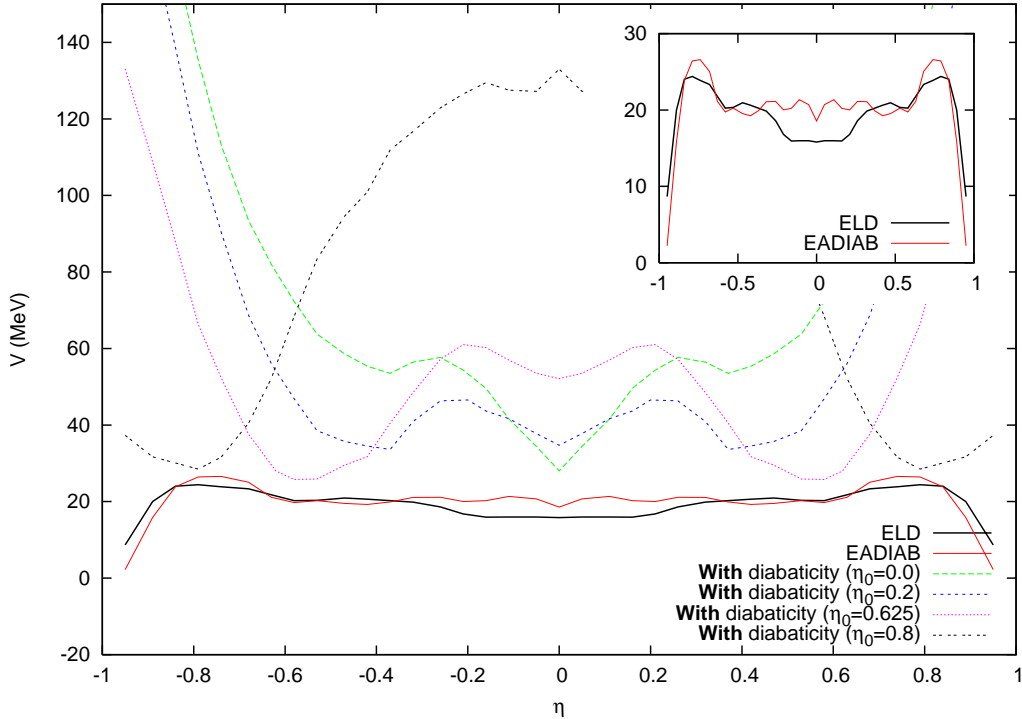}
\end{center}
\caption{(Color online) Driving potential resulted (i) from the entrance diabatic PES of Fig. 7 (curves other than the two lowest 
ones which are also shown in the small picture inserted), (ii) from the adiabatic PES at the top of Fig. 8 
(red solid curve), and (iii) from the liquid drop PES at the bottom of Fig. 8 (black solid curve). 
See text for further details.}
\end{figure}

\subsubsection{Probability distribution of the nuclear shapes and their intrinsic excitation energy}

Figs. 11-12 show the probability distribution of the nuclear shapes of the system $^{256}$No as a function of time for an entrance channel mass asymmetry $\eta_{0} = 0.0$ and 0.625, respectively. For this calculation standard values of the parameters have been used. The total incident energy and the total angular momentum are $E_{c.m.} = 30$ MeV and $J = 0\ \hbar$, respectively. It can be observed that initially during a certain period of time (few units of $10^{-22}$ s) the maximal probability remains around the contact configuration, spreading slightly in the direction of symmetric fragmentations and later on, it becomes well spread out 
over compact shapes in the fusion region. The initial diabatic PES practically relaxes into the adiabatic PES in a time about $5* 10^{-22}$ s, and before this period of time, mainly dinuclear configurations near their contact point develop and decay. Afterwards, when the repulsive core of the PES at small R no longer exists, fused compact configurations are more probable. The shell corrections to the adiabatic PES (discussed below) remain and play an important role in the final evolution of the nuclear shapes after the dissolution of the diabatic contribution to the PES. This can be seen in the structures observed in the fusion region in Figs. 11-12 (see also Fig 8).

\begin{figure}
\begin{center}
\includegraphics[width=14.0cm]{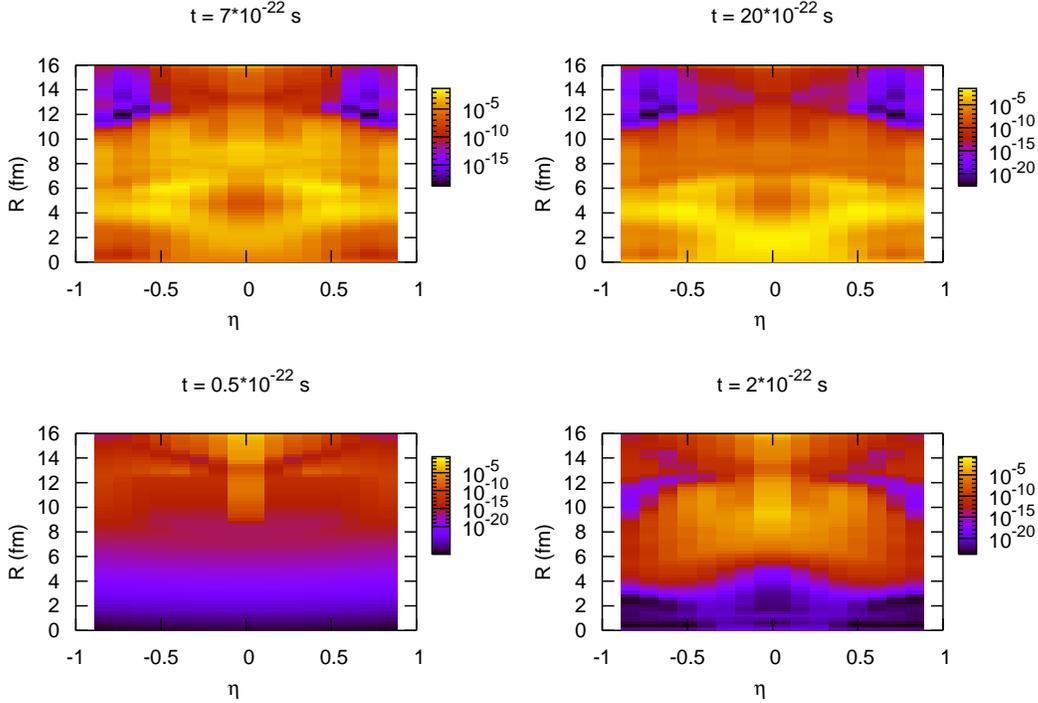}
\end{center}
\caption{(Color online) Probability distribution of the nuclear shapes as a function of time for an entrance 
channel mass asymmetry $\eta_0 = 0.0$ leading to $^{256}$No. See text for further details.}
\end{figure}

\begin{figure}
\begin{center}
\includegraphics[width=14.0cm]{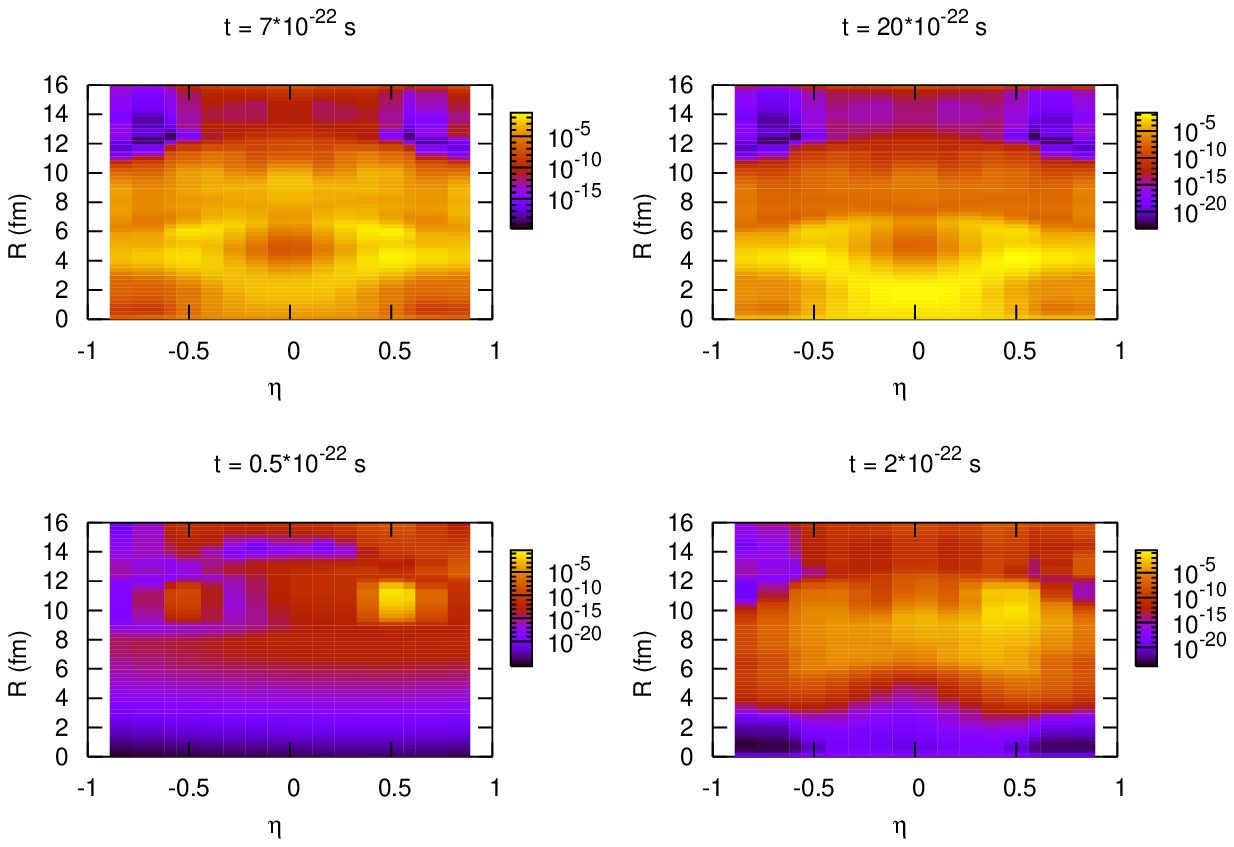}
\end{center}
\caption{(Color online) The same as Fig. 11, but for $\eta_0 = 0.625$. See text for further details.}
\end{figure}

Figs. 13-14 show the evolution of the intrinsic excitation energy of the nuclear shapes in Figs. 11-12. Very black regions are classically forbidden. Nuclear shapes there start heating up 
when these regions become classically allowed. At the quasi-fission time 
$\tau_{qf} = 2* 10^{-21}$ s the compact fused configurations are notably warmer than the quasi-fission configurations.
Therefore, the remaining ground-state shell effects on the PES are larger for the quasi-fission configurations than for the compact fused configurations. Fragmentations around the entrance channel mass asymmetry $\eta_0$ remain quite cold (quasi-elastic events), while the intrinsic excitation energy of the fragmentations far away from $\eta_0$ increases. However, some maxima can clearly be observed such as those around 
$\eta \sim \pm 0.4$ [$\epsilon_{CN}^{*}(\tau_{qf}) \sim 25$ MeV] for $\eta_{0}=0.0$ (Fig. 13) and $\eta \sim 0.0$ [$\epsilon_{CN}^{*}(\tau_{qf}) \sim 30$ MeV] for $\eta_{0}=0.625$. These structures as well as those observed in the fusion region are due to the diabatic effects and the shell corrections in the dynamical PES 
(see Figs. 8 and 10). Although the fusion-fission events are not included in the present calculations, it is shown that the quasi-fission fragments in the symmetric mass region for the entrance channel $\eta_{0}=0.625$ (very close to the system $^{48}$Ca + $^{208}$Pb) are notably much warmer (about a factor two) than those quasi-fission products with 
$\eta \sim \pm 0.4$ (around the fragmentation $^{70}$Ni + $^{186}$W) that correspond to a local maximum in the quasi-fission mass distribution discussed below. This is consistent with the experimental data of neutron and gamma multiplicity of the quasi-fission and fission products for the reaction $^{48}$Ca + $^{208}$Pb discussed by Itkis et al. in Ref. \cite{Itkis}.

In the fusion region, the intrinsic excitation energy of the nuclear shapes clearly correlates with the distribution of probability shown in Figs. 11-12. For compact fused shapes, the intrinsic excitation energy coincides with the total excitation energy. Just after fusion, the average 
intrinsic excitation energy of the CN $\epsilon_{CN}^{*}(\tau_{qf})$ is about 36.72 MeV for $\eta_{0}=0.0$ 
in Fig. 13 and about 36.76 MeV for $\eta_{0}=0.625$ in Fig. 14. 
Please note that the total incident energy $E_{c.m.}$ like the PES is normalized with respect to the LDE of the 
spherical $^{256}$No, so $\epsilon_{CN}^{*}(\tau_{qf}) > E_{c.m.}$ due to negative values of the microscopic shell 
corrections. The permanence of the shell corrections in the PES is crucial for the survival of the heaviest CN against fission. The formation of ER will be analysed in a future work.   

\begin{figure}
\begin{center}
\includegraphics[width=14.0cm]{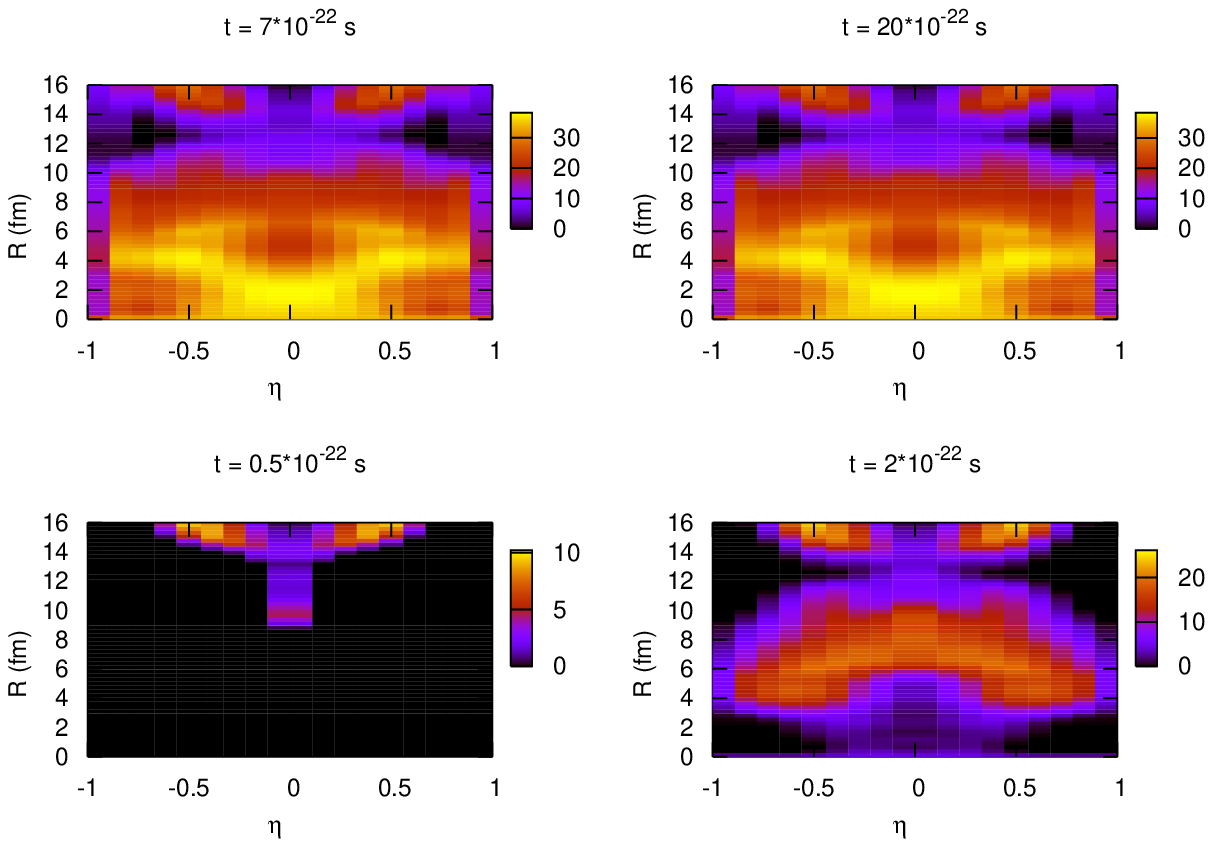}
\end{center}
\caption{(Color online) Intrinsic excitation energy (in MeV) of the nuclear shapes in Fig. 11. See text for further details.}
\end{figure}

\begin{figure}
\begin{center}
\includegraphics[width=14.0cm]{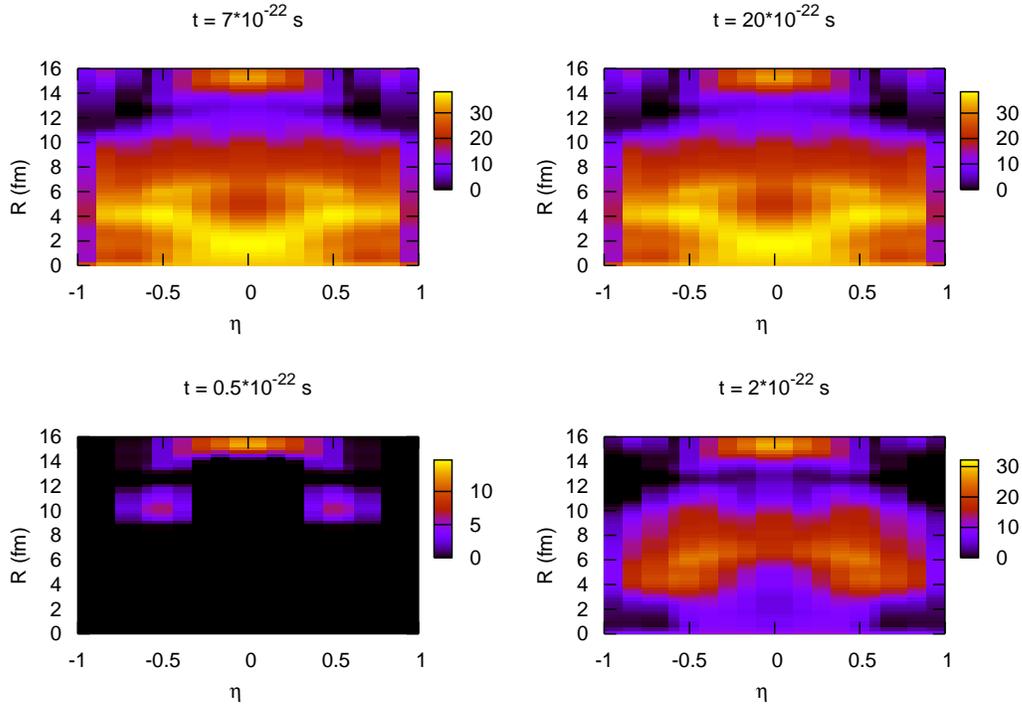}
\end{center}
\caption{(Color online) The same as Fig. 13, but the nuclear shapes are those in Fig. 12. See text for further details.}
\end{figure}
 
\subsubsection{Dependence of the observables on the parameters}

We will now concentrate on (i) the convergence of the observables with respect to the density of the mesh (i.e., $\Delta R$ and $\Delta \eta$) for fixed values of $\widetilde{\sigma}_{R}$ and $\widetilde{\sigma}_{\eta}$, (ii) the dependence of the calculation on the model critical parameters ($\Gamma_0^{-1}$ and $\kappa_0$), and (iii) the physical behaviour of the observables regarding the total incident energy $E_{c.m.}$, the total angular momentum $J$ and the entrance channel mass asymmetry $\eta_0$. At the end of the subsection we will discuss the diabatic effect as well as the effect of the shell corrections on the value of $P_{CN}$ for different entrance channels 
$\eta_0$. For the present study, several reactions leading to the $^{256}$No CN will be selected.

\textit{Mesh density}. Fig. 15 shows the dependence of some observables on $\Delta R$ (bottom) and 
$\Delta \eta$ (top) for an entrance channel mass asymmetry $\eta_0 = 0.0$ and standard values of 
the remaining parameters. The total incident energy is $E_{c.m.} = 30$ MeV, while the total angular momentum is $J = 0\ \hbar$. 

\begin{figure}
\begin{center}
\includegraphics[width=14.0cm]{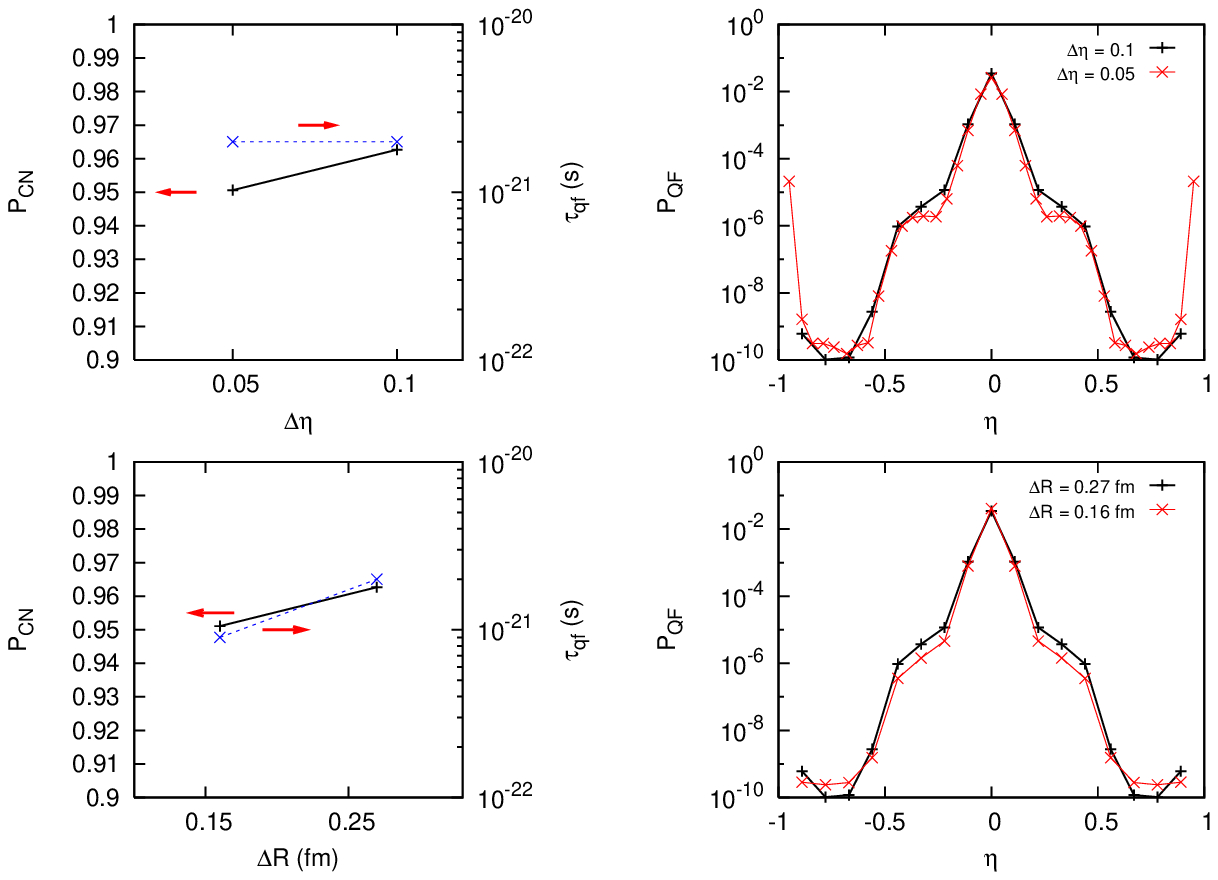}
\end{center}
\caption{(Color online) Dependence of $P_{CN}$, $\tau_{qf}$ and the mass distribution of the quasi-fission fragments 
on $\Delta R$ (bottom) and $\Delta \eta$ (top) for an entrance channel mass asymmetry $\eta_0 = 0.0$. 
See text for further details.}
\end{figure}

The results are quite stable regarding the values of $\Delta R$ (bottom) and $\Delta \eta$ (top). 
The larger the mesh density, the larger is the transition 
probability rate between two neighbouring configurations (nodes of the mesh) at fixed values of the Gaussian variances 
$\widetilde{\sigma}_{R}$ and $\widetilde{\sigma}_{\eta}$. It is reflected in a slight increase of the quasi-fission probability at large fragmentations $\eta$ and in a reduction of the 
quasi-fission time (bottom left). The large mass yield for $\eta$ very close to one (top right) is associated 
with the strong decreasing of the asymptotic adiabatic PES at very large $\eta$ (see Figs. 8 and 10). 
The decay of the system for large $R$ clearly determines the quasi-fission 
time which correlates with the $\Delta R$ values. This is because the difference between the level density 
of two neighbouring configurations is much bigger in the quasi-fission region than in the other regions due to 
the Coulomb repulsion energy. The main peak in the quasi-fission mass distribution corresponds to the entrance channel mass asymmetry $\eta_0$. The structures in the mass yield are related to the structures of the dynamical PES (see Fig. 10). These latter are initially caused by the diabatic effects, and by the remaining shell effects after the relaxation of the diabatic PES to the adiabatic one. 
The \textit{shoulders} in the quasi-fission yield around $\eta \sim \pm 0.4$ are related to the fragmentation 
$^{70}$Ni + $^{186}$W, which reflects the importance of the light fragment closed proton shell $Z = 28$. 
This is also consistent with the experiments for $^{48}$Ca + $^{208}$Pb discussed by Itkis et al. 
in Ref. \cite{Itkis}. 
 
\textit{Critical parameters}. Fig. 16 shows the dependence of the same observables discussed above on the 
parameters $\kappa_0$ (bottom) and $\Gamma_0^{-1}$ (top) for an entrance channel mass asymmetry 
$\eta_0 = 0.0$ with $E_{c.m.} = 30$ MeV and $J = 0\ \hbar$. The remaining parameters have standard values.

At the top in Fig. 16 (left) we can see that $P_{CN}$ is more sensitive to the value of 
of $\Gamma_0^{-1}$ than to the value of $\kappa_0$ (bottom left). The opposite happens with the quasi-fission 
time $\tau_{qf}$. However, taking into account that here the $P_{CN}$ scale is large, $P_{CN}$ remains rather stable in the two dependences. The smaller $\Gamma_0^{-1}$, the slower the diabatic PES relaxes. Thus more probability moves into the quasi-fission region, decreasing the value of $P_{CN}$, before the repulsive core of the diabatic PES at small $R$ disappears. In this case the quasi-fission mass distribution (red curve at the top right pannel) notably reflects the structures of the initial diabatic PES (see Fig. 10). By decreasing $\Gamma_0^{-1}$ by a factor ten, the quasi-fission time $\tau_{qf}$ increases by a factor two. It is because a significant probability still remains in the competition region after the complete relaxation of the diabatic PES to the adiabatic one.
 
\begin{figure}
\begin{center}
\includegraphics[width=14.0cm]{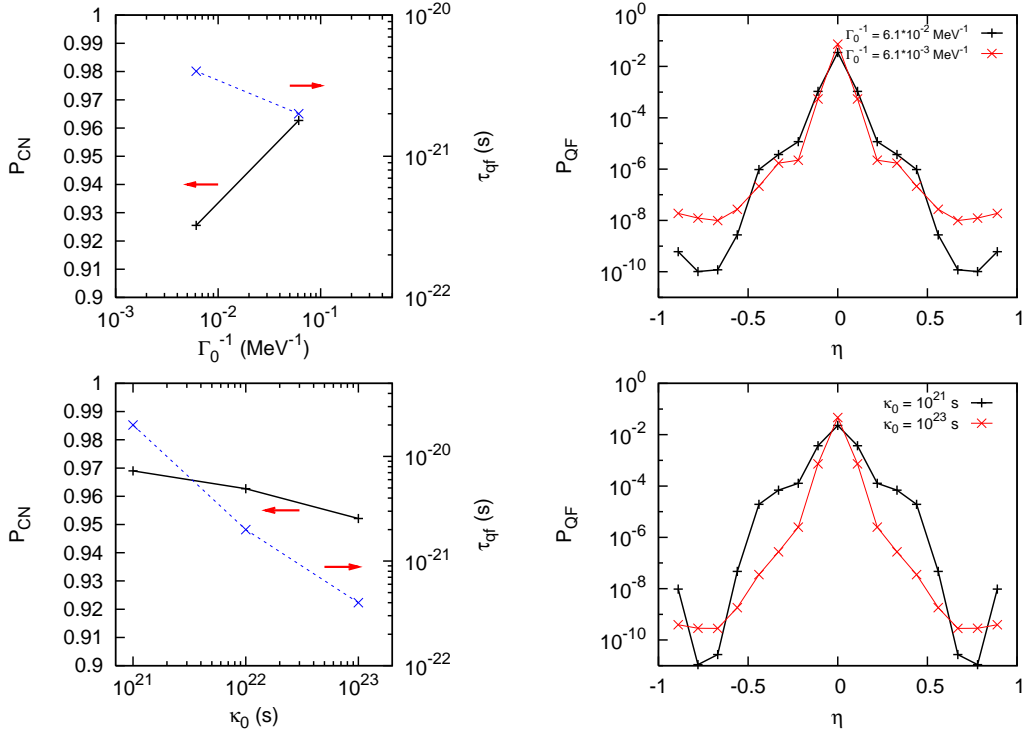}
\end{center}
\caption{(Color online) Dependence of the same observables of Fig. 15 on $\kappa_0$ (bottom) and $\Gamma_0^{-1}$ (top) for 
the entrance channel mass asymmetry $\eta_0 = 0.0$. See text for further details.}
\end{figure}

Although $P_{CN}$ very weakly depends on $\kappa_0$ (bottom left), the quasi-fission time $\tau_{qf}$ as well as the quasi-fission mass yield (bottom right) are strongly affected by the value of $\kappa_0$. 
The quasi-fission time $\tau_{qf}$ clearly correlates with the $\kappa_0$ value. The larger $\kappa_0$, the larger is the transition probability rate between the configurations, particularly in the quasi-fission region, due to the level density (phase space) effect discussed above in the dependence on $\Delta R$ values. For a given period of time (with the relaxation time of the diabatic PES being determined by the fixed $\Gamma_0^{-1}$ value) more probability moves into the quasi-fission region with increasing $\kappa_0$, thereby reducing the value of 
$P_{CN}$. The quasi-fission mass distribution (bottom right) reveals that its width decreases with 
increasing $\kappa_0$.  

\textit{Dependence on $E_{c.m.}$, $J$ and $\eta_0$}. Fig. 17 shows the dependence of the same observables 
discussed above on the total incident energy $E_{c.m.}$ (bottom, $J = 0\ \hbar$ and $\eta_0 = 0.0$), 
on the total angular momentum $J$ (middle, $E_{c.m.} = 30$ MeV and $\eta_0 = 0.0$) and on the entrance 
channel mass asymmetry $\eta_0$ (top, $J = 0\ \hbar$ and $E_{c.m.} = 30$ MeV). 
The remaining parameters have standard values.

At the bottom in Fig. 17 (left) we can observe that $P_{CN}$ reaches a maximum value 
around $E_{c.m.} = 30$ MeV, just at the Coulomb barrier of the diabatic capture potential 
(see Fig. 9, bottom left), 
while the quasi-fission time remains constant. At energies above the peak of the $P_{CN}$ excitation function, 
$P_{CN}$ slightly decreases with increasing $E_{c.m.}$. After the relaxation of the diabatic PES to the adiabatic one in a time about  
$5* 10^{-22}$ s, a significant probability still remains in the competition region. The distribution of this probability between the fusion and quasi-fission regions depends on the competition between the phase space of the 
two regions, which seems to be regulated by the incident energy $E_{c.m.}$. With decreasing $E_{c.m.}$ 
towards the capture barrier, the phase space of the fusion region seems to play a more important role and, therefore, $P_{CN}$ increases slightly. At low incident energies below the diabatic capture barrier, the fusing system spends a certain period of time in a classically forbidden region. Consequently, the phase space of the 
quasi-fission region clearly dominates and $P_{CN}$ decreases notably. 

At the bottom on the right in Fig. 17 it is shown that the mass yield of the quasi-fission fragments is correlated with the $P_{CN}$ excitation function. At the lowest incident energy, 
the mass yield shows structures that reflect those caused by the diabatic effects in the entrance diabatic PES (see Fig. 10). With increasing $E_{c.m.}$ these structures in the mass yield gradually disappear, although very smooth shoulders can still be observed at $E_{c.m.} = 80$ MeV. The quasi-fission mass distribution becomes smoother with increasing $E_{c.m.}$ reflecting the asymptotic adiabatic PES (see Fig. 10). The remainig smooth structures in the mass distribution at the highest incident energy are due to some remaining effect of the initial diabatic phase of the reaction along with the remaining ground-state shell effect of the warm quasi-fission fragments.

\begin{figure}
\begin{center}
\includegraphics[width=14.0cm]{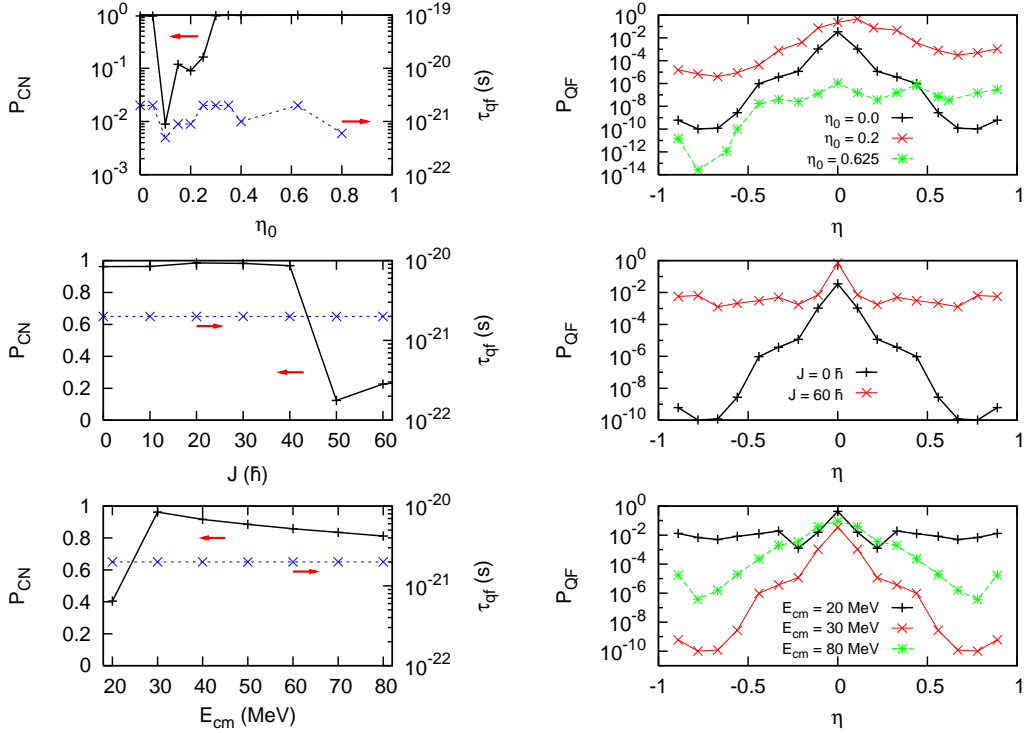}
\end{center}
\caption{(Color online) Dependence of the same observables of Fig. 15 on $E_{c.m.}$ (bottom), $J$ (middle) and $\eta_0$ (top). 
See text for further details.}
\end{figure}

In the middle in Fig. 17 (left) we see that $P_{CN}$ globally decreases with increasing $J$ and 
the quasi-fission time $\tau_{qf}$ remains constant. The values of $P_{CN}$ very weakly depend on $J$ up to $J= 40\ \hbar$ because the rotational contribution to the PES for compact shapes is small. Since the rigid-body moment of inertia, which the system reaches at the contact radius, is much larger than the moment of inertia for the orbital motion at distances around the radius of the Coulomb barrier, the centrifugal effects increase the Coulomb barrier and it becomes narrower with increasing $J$. As a consequence, $P_{CN}$ first increases slightly due to the increase of the Coulomb barrier and then (around $J= 20\ \hbar$) starts decreasing because the narrowing of the barrier clearly favors quasi-fission. For $J > 40\ \hbar$, the rotational contribution to the PES for compact shapes is so significant that these shapes are in a classically forbidden region, even after the complete relaxation of the initial diabatic PES. Thus, the dominance of the phase space of the quasi-fission region over the phase space of the fusion region is very strong. Hence, $P_{CN}$ decreases strongly. The mass yield of the quasi-fission fragments (in the middle on the right in Fig. 17) increases with increasing $J$. 

At the top in Fig. 17 (left) it is shown that $P_{CN}$ is much smaller for near-symmetric reactions compared to very asymmetric ones. This is because the quasi-fission is stronger for more symmetric entrance channels due to the effect of the diabaticity that decreases the capture valley (see Fig. 9). The capture valley becomes deeper containing more compact shapes with increasing $\eta_0$. 
On the other hand, the quasi-fission time $\tau_{qf}$ globally decreases with increasing $\eta_0$. 
Some structures are revealed, which are clearly correlated with those of the $P_{CN}$ values (see Fig. 18).
In contrast to the quasi-fission mass distribution for $\eta_0 = 0.0$ (black curve 
at the top on the right panel in Fig. 17), the mass yield of quasi-fission for $\eta_0 = 0.625$ (very close to the system 
$^{48}$Ca + $^{208}$Pb) is notably asymmetric with respect to $\eta = 0.0$. This asymmetry seems to increase with increasing $\eta_0$ (comparing the red curve for $\eta_0 = 0.2$ to the green one for $\eta_0 = 0.625$). It is important to remember that this happens in the rotating body-fixed reference frame. Further work is needed to clarify whether this anisotropy can be observed experimentally. The mass distribution for $\eta_0 = 0.625$ also shows a local minimum instead of a peak at $\eta_0$, while $P_{CN}$ reveals 
a local maximum (within $99\%$ of accuracy of the calculation) in the small picture inserted in Fig. 18. Thus a reaction induced by closed shell nuclei seems to increase $P_{CN}$.
 
\begin{figure}
\begin{center}
\includegraphics[width=14.0cm]{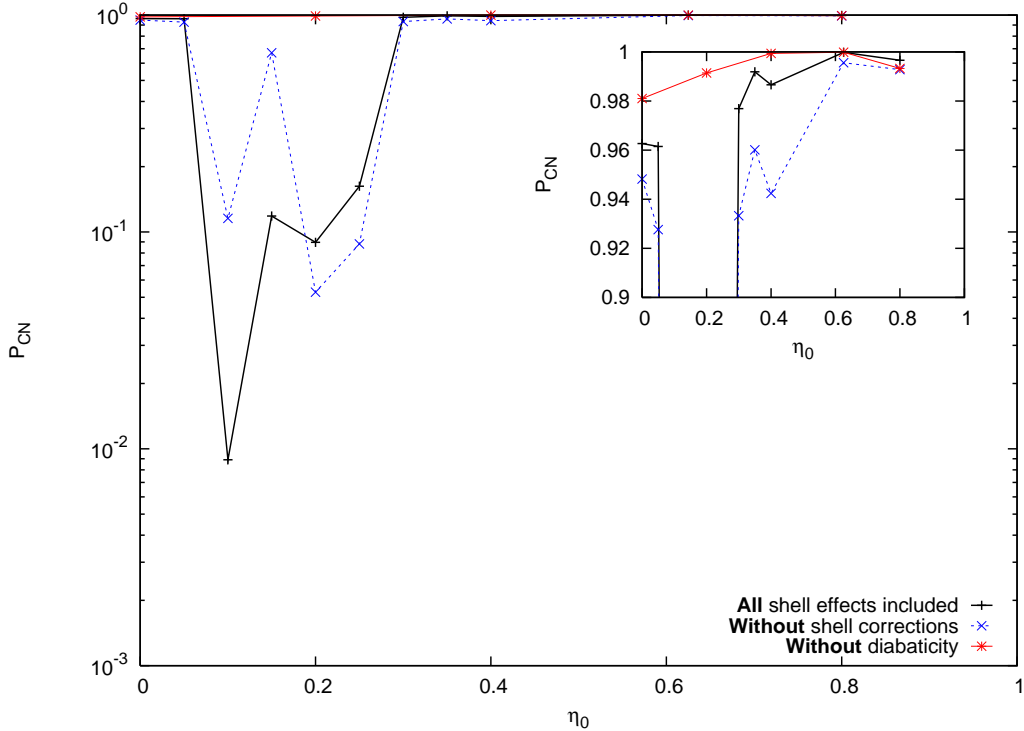}
\end{center}
\caption{(Color online) The same for $P_{CN}$ as that in the top left panel of Fig. 17 (black solid curve), but compared 
to values obtained without shell corrections (blue dashed curve) and without diabatic effects (red solid curve). See text for further details.}
\end{figure}

\textit{Diabatic and shell corrections effect}. In Fig. 18 (similar to the top left panel of Fig. 17) we can observe that the shell structure of the nuclei in the entrance channel plays an important role in establishing the value of $P_{CN}$. Here, $P_{CN}$ as a function of $\eta_0$ is shown for three types of calculation: (i) all shell effects (diabatic and shell corrections effect) are included (black solid curve), (ii) without shell corrections (blue dashed curve), and (iii) without diabatic effects (red solid curve). The effect of the diabaticity on $P_{CN}$ can be seen comparing (i) and (iii), whereas the effect of the shell corrections comparing (i) and (ii). From this study we conclude:

\begin{itemize}

\item The effect of the diabaticity in suppressing $P_{CN}$ can be very strong in near-symmetric collisions.
\item The shell corrections can play a very important role in establishing the value of $P_{CN}$. Depending on the 
entrance channel mass asymmetry, they can favor or inhibit the CN formation.
  
\end{itemize}     

The reaction with $\eta_0 = 0.0$ ($^{128}$Sb + $^{128}$Sb), although more symmetric than 
that with $\eta_0 = 0.2$, reveals a $P_{CN}$ value about 0.96 due to the favorable effect of the fragment proton number that is very near the closed proton shell $Z=50$. This value of $P_{CN}$ is similar to those for very asymmetric entrance channels. However, the symmetric entrance channel shows a large diabatic shift of the capture barrier in Fig. 9 (bottom left panel) which inhibits the capture probability. Therefore, we can expect a very asymmetric entrance channel involving closed shell nuclei to be the best suited to form a heavy compound system at a given excitation energy. 

\section{Concluding remarks and outlook}

A theoretical formulation of the competition between fusion and quasi-fission in a heavy fusing system has been presented. Fusion and quasi-fission result from a diffusion process in an ensemble 
of nuclear shapes, each of which evolves towards the thermal equilibrium. The theory is unique and more realistic than the current fusion theories because it is based on microscopical grounds and also for the first time incorporates a wide range of important physical effects that impact on the formation of the heaviest compound nuclei. The dynamical collective PES emerges as a very important new feature of this approach, which partially reconciles conflicting aspects of the current models for CN formation.

Realistic calculations for several reactions leading to $^{256}$No indicate that 
(1) the diabatic effects are very important in the capture stage of the reaction as well as in the subsequent evolution of the compact nuclear shapes in near-symmetric collisions, (2) these effects cause structures in the dynamical collective PES that are reflected in observables like the mass distribution of the quasi-fission fragments and their intrinsic excitation energy. These structures are not related to the well-known static ground-state shell corrections, but to the diabatic sp motion through the shell structure of the nuclear shapes, (3) the quasi-fission time is notably larger than the relaxation time of the initial diabatic collective PES to the adiabatic one, so the nuclear shapes reach the thermal equilibrium during the CN formation, (4) the compact fused shapes are much warmer than the quasi-fission fragments, (5) the remaining shell corrections to the adiabatic PES can play a very important role in establishing the $P_{CN}$ value, (6) $P_{CN}$ very weakly depends on the total angular momentum $J$ for small values of $J$ (e.g., up to $J \sim 40 \hbar$ for a central collision at the capture barrier energy) and reaches its maximum value around the capture barrier energy, (7) the quasi-fission mass yield increases with increasing $J$, (8) the quasi-fission time very weakly depends on the incident energy and the total angular momentum, while it globally decreases with increasing entrance channel mass asymmetry. Here, some structures depending on the entrance channel mass asymmetry are manifested, which are correlated with those of the $P_{CN}$ values, (9) the angular anisotropy of the quasi-fission fragments may increase with increasing asymmetry of the entrance channel, and (10) very asymmetric reactions induced by closed shell nuclei seem to be the best suited to synthesize the heaviest compound systems. Last but not least, the calculations are quite stable regarding their dependence on model parameters. 

The effect of other important collective degrees of freedom like the deformation of the colliding nuclei and their mutual orientations as well as the effect of the fission valley on the CN formation will also be investigated in the future. Works in this direction along with the calculation of ER cross sections are in progress.   


$\textit{Aknowledgements:}$ The author thanks C. Greiner for his support at the beginning of this project in Frankfurt, W. Scheid, D.J. Hinde and M. Dasgupta for discussions. Special thanks to the Center for Scientific Computing (CSC) at the University of Frankfurt for the computing time. 

 \appendix

\section{Derivation of the generalized master equation (\ref{5})}
 \label{Master_Equation}

The following coupled equations are obtained by acting with the
projection operators $C$ and $Q=1-C$ on the Liouville equation
(\ref{3b}) [$|\rho (t)) = C|\rho(t)) + Q|\rho(t))$ and replacing the
partial derivative by a total derivative]

\begin{equation}
i\frac{d}{dt}C|\rho (t))= CL(t)[C|\rho(t)) + Q|\rho(t))], \label{A1}
\end{equation}

\begin{equation}
i\frac{d}{dt}Q|\rho (t))= QL(t)[C|\rho(t)) + Q|\rho(t))]. \label{A2}
\end{equation}

Integrating (\ref{A2}) the solution $Q|\rho (t))$ reads

\begin{eqnarray}
Q|\rho(t))=&&\exp [-iQ\int_{t_0}^{t} dt'L(t')]\cdot
\{Q|\rho(t_0))\nonumber \\
&&- \int_{t_0}^{t} dt_1
\exp [iQ\int_{t_0}^{t_1} dt'L(t')]\ iQL(t)C|\rho (t))\}. \label{A3}
\end{eqnarray}

Inserting (\ref{A3}) into (\ref{A1}) and taking into account that
$CL(t)C=0$ \cite{Zwanzig1,Zwanzig2} because $L_{ii,ll}(t)=0$ according to the
definition (\ref{3c}), the diagonal part of the density matrix $C|\rho(t))$
obeys the following equation

\begin{eqnarray}
\frac{d}{dt}C|\rho (t))=&& -iCL(t)\exp [-iQ\int_{t_0}^{t} dt'L(t')]\cdot
Q|\rho (t_0))\nonumber \\
&&- CL(t)\int_{0}^{t-t_0} d\tau
\exp [-iQ\int_{t-\tau}^{t} dt'L(t')]QL(t-\tau)C|\rho (t-\tau)),
\nonumber \\ \label{A4}
\end{eqnarray}
where $\tau = t - t_1$. The projection on $(nn|$ and summation over
$n \mathcal{\epsilon} \mathcal{H}_{\nu}$, taking into account
expressions (\ref{4}) and (\ref{4b}), leads to

\begin{equation}
\frac{d P_{\nu} (t)}{d t} = \sum_{\mu}\int_{0}^{t-t_0} d\tau K_{\nu \mu} (t,\tau)
d_{\nu}P_{\mu} (t-\tau) + G_{\nu} (t,t_0), \label{A5}
\end{equation}
where $K_{\nu \mu} (t,\tau)$ and $G_{\nu} (t,t_0)$ are given by expressions
(\ref{6a}) and (\ref{6b}), respectively. Owing to the validity of the following relation

\begin{equation}
\sum_{\mu}d_{\mu}K_{\nu \mu}(t,\tau)=0\ \Rightarrow\
d_{\nu}K_{\nu \nu}(t,\tau)=-\sum_{\mu \neq \nu}d_{\mu}K_{\nu \mu}(t,\tau),
\label{A6}
\end{equation}
because

\begin{equation}
\sum_{\mu} \sum_{m \mathcal{\epsilon} \mathcal{H}_{\mu}}L_{ab,mm}=0,
\label{A7}
\end{equation}
according to definition (\ref{3c}), expression (\ref{A5}) turns into (\ref{5}) after
using the relation (\ref{A6}) in (\ref{A5}). The validity of (\ref{A6}) can be checked
applying the multiplication rule for tetradics \cite{Zwanzig1,Zwanzig2}
regarding the product of the terms of $K_{\nu \mu}(t,\tau)$.

\end{document}